\documentclass[prd,nofootinbib,twocolumn,preprintnumbers]{revtex4-1}
\pdfoutput=1
\usepackage{amsmath,amsthm,amssymb,multirow,psfrag}
\usepackage{epsfig}
\usepackage{color}

\graphicspath{{./Figures/}}

\begin{document}

\def\lsim{\mathrel{\rlap{\lower4pt\hbox{\hskip1pt$\sim$}}
\raise1pt\hbox{$<$}}}
\def\gsim{\mathrel{\rlap{\lower4pt\hbox{\hskip1pt$\sim$}}
\raise1pt\hbox{$>$}}}
\newcommand{\vev}[1]{ \left\langle {#1} \right\rangle }
\newcommand{\bra}[1]{ \langle {#1} | }
\newcommand{\ket}[1]{ | {#1} \rangle }
\newcommand{\ev}{ {\rm eV} }
\newcommand{\kev}{{\rm keV}}
\newcommand{\mev}{{\rm MeV}}
\newcommand{\gev}{{\rm GeV}}
\newcommand{\tev}{{\rm TeV}}
\newcommand{\mpl}{$M_{Pl}$}
\newcommand{\mw}{$M_{W}$}
\newcommand{\Ft}{F_{T}}
\newcommand{\Zparity}{\mathbb{Z}_2}
\newcommand{\BLambda}{\boldsymbol{\lambda}}
\newcommand{\bea}{\begin{eqnarray}}
\newcommand{\eea}{\end{eqnarray}}
\newcommand{\bwt}{\begin{widetext}}
\newcommand{\ewt}{\end{widetext}}
\newcommand{\met}{\;\not\!\!\!{E}_T}
\newcommand{\draftnote}[1]{{\bf\color{blue} #1}}
\newcommand{\draftnoteR}[1]{{\bf\color{red} #1}}
\newcommand{\draftnoteG}[1]{{\bf\color{green} #1}}

\newcommand{\fref}[1]{Fig.~\ref{fig:#1}} 
\newcommand{\eref}[1]{Eq.\eqref{eqn:#1}} 
\newcommand{\aref}[1]{Appendix~\ref{app:#1}}
\newcommand{\sref}[1]{Sec.~\ref{sec:#1}}
\newcommand{\ssref}[1]{Sec.~\ref{subsec:#1}}
\newcommand{\tref}[1]{Table~\ref{tab:#1}}

\title{{\bf 8D Likelihood  Effective Higgs Couplings Extraction Framework in $h\to 4\ell$}}

\author{Yi Chen}
\email{yi.chen@cern.ch}
\affiliation{Lauritsen Laboratory of Physics, California Institute of Technology, 1200 E. California Blvd, Pasadena, CA, 92115, USA}

\author{Emanuele Di Marco}
\email{emanuele.di.marco@cern.ch}
\affiliation{Lauritsen Laboratory of Physics, California Institute of Technology, 1200 E. California Blvd, Pasadena, CA, 92115, USA}

\author{Joe Lykken}
\email{lykken@fnal.gov}
\affiliation{Theoretical Physics Department, Fermilab, P.O.~Box 500, Batavia, IL 60510, USA}

\author{Maria Spiropulu}
\email{smaria@caltech.edu}
\affiliation{Lauritsen Laboratory of Physics, California Institute of Technology, 1200 E. California Blvd, Pasadena, CA, 92115, USA}

\author{Roberto Vega-Morales}
\email{roberto.vega@th.u-psud.fr}
\affiliation{Laboratoire de Physique Th\'{e}orique d'Orsay, UMR8627-CNRS, Universit\'{e} Paris-Sud 11, F-91405 Orsay Cedex, France}
\affiliation{Theoretical Physics Department, Fermilab, P.O.~Box 500, Batavia, IL 60510, USA}
\affiliation{Department of Physics and Astronomy, Northwestern University, 2145 Sheridan Road, Evanston, IL 60208, USA}

\author{Si Xie}
\email{sixie@hep.caltech.edu}
\affiliation{Lauritsen Laboratory of Physics, California Institute of Technology, Pasadena, CA, 92115}


\begin{abstract}
We present an overview of a comprehensive analysis framework aimed at performing direct extraction of all possible effective Higgs couplings to neutral electroweak gauge bosons in the decay to electrons and muons, the so called `golden channel'.~Our framework is based primarily on a maximum likelihood method constructed from analytic expressions of the fully differential cross sections for $h \rightarrow 4\ell$ and for the dominant irreducible $q\bar{q} \rightarrow 4\ell$ background, where $4\ell = 2e2\mu, 4e, 4\mu$.~Detector effects are included by an explicit convolution of these analytic expressions with the appropriate transfer function over all center of mass variables.~Utilizing the full set of observables, we construct an un-binned detector-level likelihood which is continuous in the effective couplings.~We consider possible $ZZ$, $Z\gamma$, and $\gamma\gamma$ couplings simultaneously, allowing for general CP odd/even admixtures.~A broad overview is given of how the convolution is performed and we discuss the principles and theoretical basis of the framework.~This framework can be used in a variety of ways to study Higgs couplings in the golden channel using data obtained at the LHC and other future colliders. 
\end{abstract}
\preprint{LPT-Orsay-13-100, CALT-68-2874, FERMILAB-PUB-13-555-T, nuhep-th/13-06}

\maketitle


\section{Introduction}
\label{sec:Intro}

The recent discovery of the Higgs boson at the LHC~\cite{:2012gk,:2012gu} with properties resembling those predicted by the Standard Model, shifts our attention to the determination of its precise nature and to establish whether or not the Higgs boson possesses any anomalous couplings to Standard Model particles.~In this study we focus on couplings to neutral electroweak gauge bosons.~Since these `anomalous effects' are expected to be small if at all present, constraining or measuring of these couplings should preferably be done through direct parameter extraction with minimal theoretical assumptions.~The vast literature~\cite{Nelson:1986ki,Soni:1993jc,Chang:1993jy,Barger:1993wt,Arens:1994wd,Choi:2002jk,Buszello:2002uu,Godbole:2007cn,Kovalchuk:2008zz,Cao:2009ah,Gao:2010qx,DeRujula:2010ys,Gainer:2011xz,Coleppa:2012eh,Bolognesi:2012mm,Stolarski:2012ps,Chen:2012jy,Boughezal:2012tz,Belyaev:2012qa,Avery:2012um,Campbell:2012ct,Campbell:2012cz,Chatrchyan:2012sn,Chatrchyan:2012ufa,Chatrchyan:2012jja,Modak:2013sb,Sun:2013yra,Gainer:2013rxa,Artoisenet:2013puc,Anderson:2013fba,Chen:2013waa,Buchalla:2013mpa,Chen:2013ejz,Gainer:2014hha,Chen:2014gka} on Higgs decays to four charged leptons (electrons and muons) through neutral electroweak gauge bosons, suggests that the so called `golden channel', can be a powerful means towards accomplishing this goal.

A number of frameworks have been established utilizing the Matrix Element Method to study the golden channel aiming to determine these potentially anomalous couplings.~These primarily rely on Monte Carlo generators such as the JHU generator~\cite{Gao:2010qx,Bolognesi:2012mm,Anderson:2013fba} or on Madgraph implementations~\cite{Avery:2012um,Artoisenet:2013puc}.~They have the advantage of flexibility to include various Higgs production and decay channels and are especially useful for constructing kinematic discriminators to distinguish between competing hypotheses.

Focusing on the golden channel only\footnote{Though we will not discuss it explicitly here, we are also able to extend our framework to the $h \to \gamma\gamma$ and $h \to 2\ell\gamma$ channels.}, we propose a novel analysis framework largely based on an analytic implementation.~It is designed to maximize the information contained in each event with the aim of direct extraction of the various effective Higgs couplings.~It is generally acknowledged in the literature that analytic methods are optimal for performing this direct multi-parameter extraction within practical and reasonable computational processing resources~\cite{Gao:2010qx,Bolognesi:2012mm,Anderson:2013fba}.~In this work, we also demonstrate that within an analytic framework one can readily include the relevant detector effects and obtain a detector-level likelihood function in terms of the full set of observables available in the four lepton final state.~This is accomplished by the explicit convolution of analytic expressions for the `truth level' fully differential cross sections with a transfer function which parametrizes the detector resolution and acceptance effects.

This analysis framework has already proved useful in constraining effective Higgs couplings as demonstrated in a recent CMS analysis~\cite{CMS-PAS-HIG-14-014,Khachatryan:2014kca}.~It was shown that for simplified cases of constraining one or two parameters, our framework gives comparable performance to other established analysis methods for the golden channel~\cite{Gao:2010qx,Bolognesi:2012mm,Avery:2012um,Anderson:2013fba,Artoisenet:2013puc,CMS-PAS-HIG-14-014,Khachatryan:2014kca}.~In this work we present an overview of the framework and discuss the principles and theoretical basis.~In particular we sketch how the various components of the detector level likelihood are constructed with emphasis on how the convolution integral is performed as well as various validations.~How the likelihood can then be used to perform multi-parameter extraction of effective Higgs couplings is also discussed.~We hope that the additional features of our framework are also found useful in the next phase of the LHC and future colliders.~Much more information on the framework including technical details can be found in~\cite{Chen:2012jy,Chen:2013ejz,Chen:2014gka,CMS-PAS-HIG-14-014,Khachatryan:2014kca,Chen:2014hqs,Thesis}.

\section{Overview of Framework}
\label{sec:overview}

Though `truth' level (or generator) studies of $h\to 4\ell$ ($4\ell = 2e2\mu, 4e, 4\mu$) give a good approximate estimate of the expected sensitivity to the Higgs $ZZ$, $Z\gamma$, and $\gamma\gamma$ couplings~\cite{Chen:2014gka}, when analyzing data obtained at the LHC (or future colliders) a detector level likelihood which accounts for the various detector effects is necessary.~Since generally detector level likelihoods are obtained via the use of Monte Carlo methods, it becomes difficult to obtain the full multi-dimensional likelihood for the $4\ell$ final state.~Typically one needs to fill large multi-dimensional templates that require an impractical amount of computing time.~There are also potential collateral binning and `smoothing' side-effects often associated with these methods.~In the case of the golden channel this necessitates the use of kinematic discriminants which `collapse' the fully multi-dimensional likelihood into two or perhaps three detector level observables~\cite{Anderson:2013fba}.~This approach is normally taken to facilitate the inclusion of detector effects, but is not optimal when fitting to a large number of parameters simultaneously~\cite{Bolognesi:2012mm,gaotalk}.~This is unfortunate in the case of the golden channel where in principle there are twelve observables which can be used to extract a large number of parameters at once, including their correlations.~It would be satisfying and useful to have a framework which is free of these issues and capable of utilizing all available information in the four lepton final state at detector level.

\subsection{From `Truth' to `Detector' Level}
\label{subsec:overview2}

This is accomplished in our framework by performing an explicit convolution of the generator (`truth') level probability density, formed out of the signal and background differential cross sections, with a transfer function which encapsulates the relevant detector effects.~This can be represented schematically as follows,
\bea
\label{eqn:conv}
&& P(\vec{X}^\mathrm{R} | \vec{\mathcal{A}} ) = \int P(\vec{X}^\mathrm{G} | \vec{\mathcal{A}} ) T({\vec{X}^{R} | \vec{X}^{G}}) d\vec{X}^\mathrm{G}.
\eea
Here we take $\vec{X}$ to represent the full set of center of mass variables, of which there are twelve in the golden channel, to be discussed more below, and $\vec{\mathcal{A}}$ represents some set of lagrangian parameters.~The transfer function~$T({\vec{X}^{R} | \vec{X}^{G}})$ takes us from generator (G) level to reconstructed (R) level observables and represents the probability of reconstructing the observables $\vec{X}^R$ given the generator level observable $\vec{X}^G$.~It is treated as a function of $\vec{X}^R$ which takes $\vec{X}^G$ as input.~As will be described more in~\ssref{pdf_norm}, once the integration in~\eref{conv} is performed we must then normalize over all twelve reconstructed level observables to obtain the detector level \emph{pdf}.

The integral in~\eref{conv} is the defining feature of our framework and has been obtained for both the $h\to4\ell$ signal as well as the dominant $q\bar{q}\to4\ell$ background, which have been computed analytically in accompanying studies~\cite{Chen:2012jy,Chen:2013ejz,Vega-Morales:2013dda}.~We emphasize that the integral has not been obtained via Monte Carlo methods.~Instead we have explicitly performed the integration by utilizing various analytic and well-established numerical methods~\cite{Bradie:943117,Thesis} (for studies that perform similar convolutions using Monte Carlo methods see~\cite{Artoisenet:2008zz,Artoisenet:2010cn,Artoisenet:2013puc,Artoisenet:2013vfa}).~This ensures that (arbitrarily) high precision is maintained at each step, producing what is effectively an `analytic function' in terms of detector level variables once the convolution has been performed.~After performing this 12-dimensional integration and normalizing, we are left with a probability density function (\emph{pdf}) from which \emph{we construct an un-binned twelve-dimensional detector level likelihood} which is a continuous function of the effective couplings (or Lagrangian parameters) and takes as its input, up to twelve reconstructed (detector-level) center of mass observables.~In the current implementation we will average over the four production variables to reduce the systematic uncertainties, thus obtaining an eight-dimensional likelihood in terms of just decay observables.~However, this step is in principle not necessary.

We also emphasize that the convolution integral is largely independent of detector transfer function and generator level differential cross section and in particular how accurate the descriptions of the `truth' level \emph{pdfs} or the detector properties are.~This means the framework can in principle be adapted to any detector which studies $h \to 4\ell$ (or any $X \to 4\ell$) and, in addition, as theoretical calculations of the generator level differential cross sections improve they can easily be incorporated into the convolution integral.~Thus, there is ample for room optimization in our framework as time goes on.~The generality of the convolution also allows for other beyond the Standard Model physics such as exotic Higgs decays~\cite{Falkowski:2014ffa} to be easily be incorporated into the $h\to 4\ell$ framework.

\subsection{Analytic Parameterizations}
\label{subsec:overview3}

As we discuss below, it is essential to first have analytic parameterizations of the `truth' level differential cross sections in order to perform the convolution integral in~\eref{conv}.~These can be obtained in essentially two different ways.~The first is to simply analytically compute the differential cross section starting from Feynman diagrams, which of course is not always possible.~The second is to obtain an `analytic' parameterization by fitting to a large Monte Carlo sample with some appropriately parametrized function.~This becomes quite difficult when the function is multi-dimensional as is the case in the golden channel with twelve center of mass observables and requires large samples and an accurate interpolation procedure.~In this framework we have implemented a hybrid of these two approaches with the primary component coming from analytic expressions of the leading order $h \rightarrow 4\ell$ and $q\bar{q} \rightarrow 4\ell$ fully differential cross sections~\cite{Chen:2012jy,Chen:2013ejz,Vega-Morales:2013dda}.

Since NLO effects in the golden channel are generally small~\cite{Bredenstein:2006rh,Bredenstein:2006nk,Adam:2008aa}, these leading order differential cross sections represent the dominant contributions to the $4\ell$ `truth' level likelihood.~There are however, a number of sub-dominant effects which appear at higher order and should be accounted for.~These include production and additional background effects.~In these cases, the second method of parametric fits to simulated data is typically the optimal route.~To do this we follow a similar procedure as found in~\cite{Anderson:2013fba} while further details on the implementation into our framework can be found in~\cite{Thesis}.~We emphasize however that the convolution integral is independent of these matters allowing for easy implementation of more precise `truth' level likelihoods as they become available over time.

Of course when considering the detector level likelihood there are additional, but again sub-dominant, effects not present at `truth' level which should be accounted for, such as detector momentum resolution and acceptance effects.~These can be parametrized via transfer functions which can be optimized for a particular detector as done recently in~\cite{CMS-PAS-HIG-14-014,Khachatryan:2014kca} which incorporates a parameterization of the CMS detector into our framework.~Since these typically would be supplied by the experimentalist we do not discuss their construction in detail here, but note that as knowledge of the detectors improves and parameterizations of the transfer functions become more accurate, they can easily be incorporated into the convolution integral in~\eref{conv}, but again the integration is independent of these matters.~More details on the construction and implementation, as well as the validation, of the transfer functions is found in~\cite{CMS-PAS-HIG-14-014,Khachatryan:2014kca,Thesis}.

\subsection{Fast Parameter Extraction}
\label{subsec:overview4}

The convolution integral in~\eref{conv} allows us to (effectively) obtain an analytic function in both detector level observables and lagrangian parameters.~This is because, via the explicit 12-dimensional integration over all center of mass variables, we are able to obtain a 1-to-1 mapping from the `truth' level likelihood to the `detector' level likelihood.~This allows us, during parameter extraction, to effectively work directly with the lagrangian parameters, but at detector level which gives us the ability to easily perform multi-parameter extraction with the same speed and flexibility as was done at generator level~\cite{Chen:2013ejz,Chen:2014gka}.~Being able to fit to multiple parameters simultaneously is important since it allows for strong tests of models which often predict correlations between the various parameters.~We point out that our framework allows us to do this while avoiding relying on hypothesis testing or on the construction of kinematic discriminants which is less optimal when extracting multiple parameters than maximizing the full likelihood~\cite{Neyman} where all observables are used.~Furthermore, the analytic nature of our framework allows for a great deal of flexibility in performing a variety of types of parameter extractions and re-parameterizations.

\subsection{Comments on Assumptions\\~~~~~~and Approximations}
\label{subsec:overview5}

In performing the convolution integral in~\eref{conv} we have relied on two key assumptions.~The first is that angular resolution effects due to detector smearing can be neglected,~which is an excellent approximation for the LHC detectors~\cite{CMSperformance2,CMSperformance,CMS-DP-2013-003}.~Second, we have assumed in the transfer function that each lepton is independent of the others which again is a very good approximation since leptons are clean and well-measured objects in the CMS and ATLAS detectors once standard lepton selection criteria are imposed~\cite{CMSperformance2,CMSperformance,CMS-DP-2013-003}.~With these simplifying assumptions the convolution integral can then be performed as will be described below and in much more detail in~\cite{Chen:2014hqs} and~\cite{Thesis}.

Even after the convolution is performed however, we must still normalize the detector level differential cross section.~Since this can not be done analytically one must resort to Monte Carlo techniques.~Thus, strictly speaking the final \emph{pdf} is not analytic.~However, as we will discuss more in~\ssref{pdf_norm}, due to the manner in which the analytic expressions are organized, a high precision on the normalization can be obtained in a short amount of computing time.~Furthermore, by fitting to ratios of couplings, we can circumvent the need for the \emph{absolute} normalization which greatly simplifies the computational procedure and allows us to achieve a high precision~\cite{Chen:2014hqs,Thesis} leading in the end to a detector level \emph{pdf} which is \emph{effectively analytic} in reconstructed observables and lagrangian parameters.

Of course there are components of both the `truth' and detector level likelihoods which can not be included in the convolution integral of~\eref{conv}.~These correspond to any components for which a sufficiently accurate analytic parameterization can not be obtained.~These may include potential higher order contributions to both signal and background differential cross sections as well as additional fake backgrounds such as $Z+X$.~For these one must resort to more conventional Monte Carlo techniques and the construction of large (binned) `look-up' tables.~The effects of binning can me mitigated through a linear multi-dimensional interpolation technique which is described in more detail in~\cite{Thesis}.~Fortunately these components are sub dominant in the golden channel and can be assigned systematics to study their effects~\cite{CMS-PAS-HIG-14-014,Khachatryan:2014kca,Thesis}.

There are also various other systematics associated with both detector and theoretical uncertainties which should properly be accounted for.~Since these components do not have a large effect on the final sensitivity (especially once sizable data sets are accumulated) and are not directly related to the convolution, we will discuss them only briefly below, but see~\cite{CMS-PAS-HIG-14-014,Khachatryan:2014kca,Thesis} for more details on how they are implemented into the framework in a real experimental analysis.

In constructing the detector level likelihood we have overcome many of the technical challenges which in the past have made it impossible to use the fully multi-dimensional likelihood during parameter fitting.~Below we sketch in more detail how these various challenges have been overcome, but many of the details are technically beyond the scope of this paper so we refer the reader to~\cite{Chen:2012jy,Chen:2013ejz,Chen:2014gka,CMS-PAS-HIG-14-014,Khachatryan:2014kca,Chen:2014hqs,Thesis} for more details.

\section{The `Truth Level' \emph{pdf}}
\label{sec:construct_gen_pdf} 

Before obtaining the detector level likelihood one must of course first construct the `truth' level (or generator level) likelihood.~As we discuss, the generator level likelihood is composed of a `decay' and `production' differential spectrum.~In our framework, the primary component is constructed out of analytic expressions for the $h \rightarrow 4\ell$ signal and the dominant $q\bar{q} \rightarrow 4\ell$ background differential cross sections.~Analytic expressions have been shown to be useful in likelihood methods where the full kinematics of an event can be exploited.~This is especially true for the golden channel as has been demonstrated in numerous studies~\cite{DeRujula:2010ys, Bolognesi:2012mm, Gao:2010qx, Gainer:2011xz, Stolarski:2012ps,Sun:2013yra,Anderson:2013fba,Chen:2013ejz,Chen:2014gka}.~For a detailed description of the analytic calculations for the signal and background fully differential cross sections as well as their validation we refer the reader to accompanying studies~\cite{Chen:2012jy,Chen:2013ejz}.

Below we give an overview of how the `truth' level likelihood is constructed and define the twelve center of mass variables in the four lepton final state.~We briefly discuss our parameterization of the Higgs couplings to electroweak gauge bosons and how the analytic expressions are combined with the appropriate production spectra to form the full truth level differential cross section.~We also discuss in this section how the production spectrum is obtained and comment on the additional backgrounds present in the golden channel.

\subsection{Center of Mass Observables}
\label{subsec:events}

\begin{figure}
\centering
\includegraphics[width=0.48\textwidth]{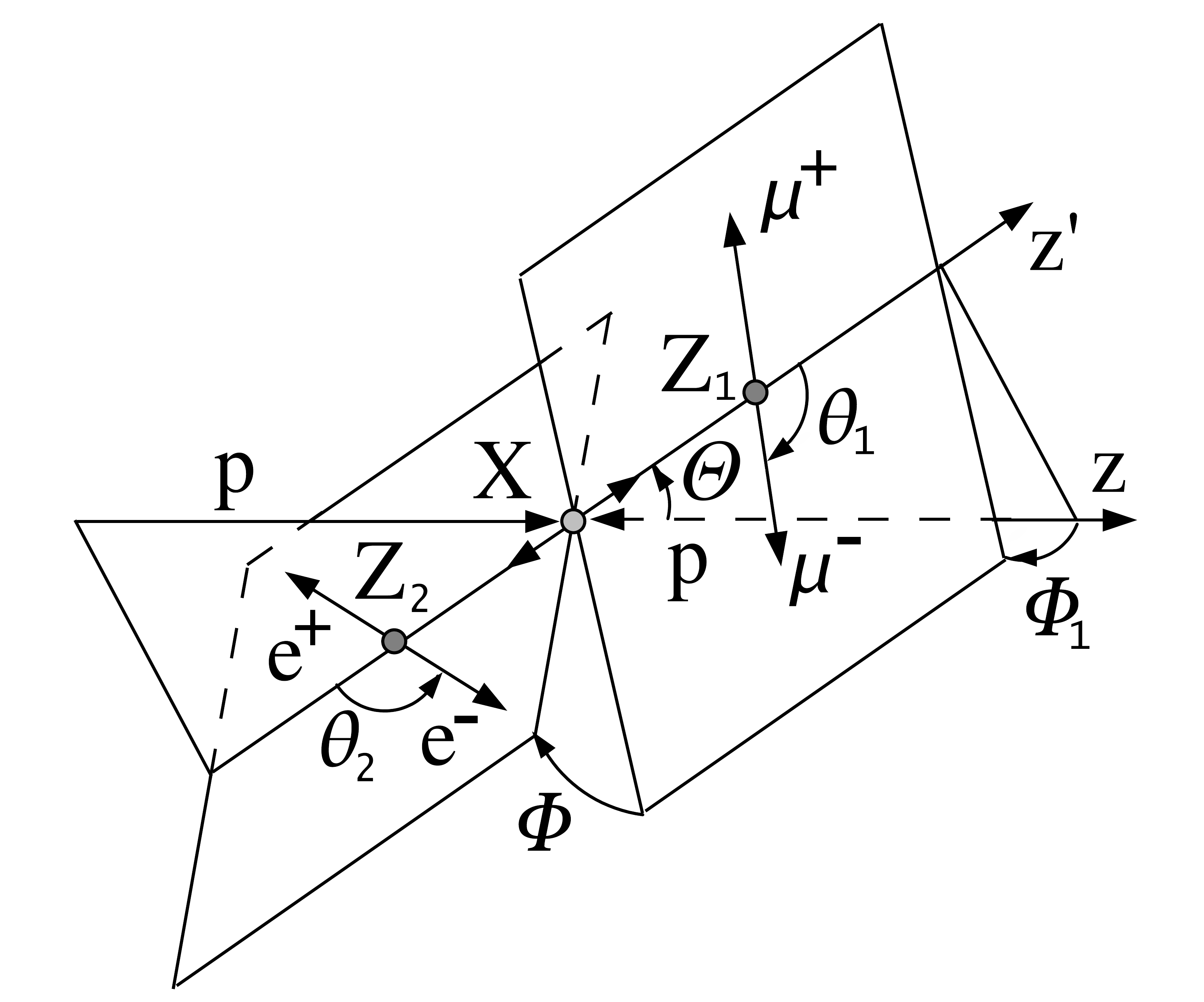}
\caption{Definition of angles in the four lepton center of mass frame $X$.}
\label{fig:DecayPlanes}
\end{figure}

Here we describe the various center of mass variables which will be used as our set of observables when constructing the likelihood.~The kinematics of four lepton events are illustrated in Fig.~\ref{fig:DecayPlanes}.
The invariant masses are defined as the following:
\begin{itemize}
\item $\sqrt{\hat{s}} \equiv M_{4\ell} \equiv m_h$ -- The invariant mass of the four lepton system or the Higgs mass in case of signal. 
\item $M_{1}$ -- The invariant mass of the lepton pair system which reconstructs closest to the $Z$ mass. 
\item $M_{2}$ -- The invariant mass of the other lepton pair system and interpreted as $M_2 < M_1$.~This condition holds as long as $\sqrt{\hat{s}} \lesssim 2 m_Z$.
\end{itemize}
These invariant masses are all independent subject to the constraint $(M_1 + M_2) \leq \sqrt{\hat{s}}$ and serve as the most strongly discriminating observables between different signal hypothesis as well as between signal and background.~Note also that the $4e/4\mu$ final state can be reconstructed in two different ways due to the identical final state interference.~This is a quantum mechanical effect that occurs at the amplitude level and thus both reconstructions are valid.~The definitions $M_1$ and $M_2$ remain unchanged however.

The angular variables are defined as:
\begin{itemize}
\item $\Theta$ -- The production angle between the momentum vectors of the lepton pair which reconstructs to $M_1$ and the total $4\ell$ system momentum.
\item $\theta_{1,2}$ -- Polar angle of the momentum vectors of $e^-,\mu^-$ in the lepton pair rest frame.
\item $\Phi_1$ -- The angle between the plane formed by the $M_1$ lepton pair and the `production plane' formed out of the momenta of the incoming partons and the momenta of the two lepton pair systems.
\item $\Phi$ -- The angle between the decay planes of the final state lepton pairs in the rest frame of the $4\ell$ system.
\end{itemize}
We group the angular variables as follows $\vec{\Omega} = (\Theta, \cos\theta_1, \cos\theta_2, \Phi_1, \Phi)$.~These angular variables are useful in aiding to distinguish different signal hypothesis and in particular between those with different CP properties, as well as in discriminating signal from background.
There are also additional production variables associated with the initial partonic state four momentum:
\begin{itemize}
\item $\vec{p}_T$ -- The momentum in the transverse direction.
\item $Y$ -- Defined as the motion along the longitudinal direction.
\item $\phi$ -- Defines a global rotation of the event in the $4\ell$ rest frame and in general does not aid greatly in discriminating power.
\end{itemize}
Including $Y$ and $\vec{p}_T$ as observables in the likelihood increases the discriminating power of the golden channel.~However, including these production variables can introduce large uncertainties since their spectra includes parton distribution functions as well as NLO contributions which should be included.~Thus, they are often integrated out of the final likelihood~\cite{CMS-PAS-HIG-14-014,Khachatryan:2014kca} and not used during parameter extraction.~This reduces the potential discriminating power, but this is compensated by the smaller systematic uncertainties one obtains by not including them in the likelihood (or averaging over them).~We will discuss more below how in our framework one can easily either include them in the final likelihood or average over them to mitigate the effects of the uncertainties associated with these variables.~However as with $\phi$, it is crucial to include them when performing the convolution with the transfer function in order to obtain the proper detector level likelihood.~These variables exhaust the twelve possible center of mass observables available in the golden channel.

\subsection{Parameterization of Scalar-Tensor Couplings}
\label{subsec:Tensor_Couplings}

Assuming only Lorentz invariance, the general couplings of a spin-0 particle to two spin-1 vector bosons can be parametrized in terms of effective couplings by the following tensor structure, 
\begin{widetext}
\bea
\label{eqn:vertex}
\Gamma^{\mu\nu}_{i} &=& \frac{i}{v} 
\left(
\mathcal{A}_1^{i} m_Z^2 g^{\mu\nu} 
+ \mathcal{A}_2^{i} (k_1^\nu k_2^\mu - k_1\cdot k_2 g^{\mu\nu}) 
+ \mathcal{A}_3^{i} \epsilon^{\mu\nu\alpha\beta} k_{1\alpha} k_{2\beta}
+ \Big( \mathcal{A}_4^{i} (\frac{k_1^2 + k_2^2}{m_Z^2})
+  \mathcal{A}_5^{i} (\frac{\hat{s}}{m_h^2}) \Big) m_Z^2 g^{\mu\nu} 
\right) ,
\eea
\end{widetext}
where in the golden channel $i=ZZ,Z\gamma,\gamma\gamma$.~The variables $k_1$ and $k_2$ represent the four momentum of the intermediate vector bosons with $v$ the Higgs vacuum expectation value (vev) which we have chosen as our overall normalization.~The $\mathcal{A}_{n}^{i}$ are dimensionless and in principle arbitrary complex form factors with possible momentum dependence (or more precisely a $\hat{s}, k_1^2, k_2^2$ dependence)~making~\eref{vertex} completely general.~Note that the tensor structure for $\mathcal{A}_5^{i}$ is only distinguishable from $\mathcal{A}_1^{i}$ for off-shell Higgs decays as discussed in~\cite{Gainer:2014hha}.~For a purely Standard Model Higgs we have $\mathcal{A}_{1}^{ZZ} = 2$ at tree level while all other effective couplings are generated at higher loop order and at most $\mathcal{O}(\lesssim 10^{-2})$.

Of course it is often possible to expand the $\mathcal{A}_{n}^{i}$ in a power series of momenta keeping only the leading (constant) terms.~By keeping the leading terms in this expansion there is a one-to-one mapping from this vertex onto the effective Lagrangian\footnote{This lagrangian has been implemented~\cite{WEBSITE} into the FeynRules/Madgraph~\cite{Christensen:2008py,Alwall:2007st} framework for validation purposes.},
\bwt
\bea\label{eqn:siglag}
\mathcal{L} \supset \frac{1}{4v}
\Big(
2 \mathcal{A}_{1o}^{ZZ} m_Z^2 h Z^\mu Z_\mu 
&+& \mathcal{A}_{2o}^{ZZ} h Z^{\mu\nu}Z_{\mu\nu} 
+ \mathcal{A}_{3o}^{ZZ} h Z^{\mu\nu} \widetilde{Z}_{\mu\nu} 
- 4 \mathcal{A}_{4o}^{ZZ} h Z_\mu\Box Z^\mu 
- 2 \mathcal{A}_{5o}^{ZZ} (\frac{m_Z}{m_h})^2 \Box h Z_\mu Z^\mu \nonumber \\
&+&
2\mathcal{A}_{2o}^{Z\gamma} h F^{\mu\nu}Z_{\mu\nu} 
+ 2\mathcal{A}_{3o}^{Z\gamma} h F^{\mu\nu} \widetilde{Z}_{\mu\nu}
+
\mathcal{A}_{2o}^{\gamma\gamma} h F^{\mu\nu}F_{\mu\nu} 
+ \mathcal{A}_{3o}^{\gamma\gamma} h F^{\mu\nu} \widetilde{F}_{\mu\nu} 
\Big) ,
\eea
\ewt
where the $\mathcal{A}_{no}^i$ represent the leading, momentum independent coefficients and electromagnetic gauge invariance requires $\mathcal{A}_{1o}^{Z\gamma,\gamma\gamma} = \mathcal{A}_{4o}^{Z\gamma,\gamma\gamma} = \mathcal{A}_{5o}^{Z\gamma,\gamma\gamma} = 0$.~We have defined $Z^\mu$ and $A^\mu$ as the $Z$ and photon fields respectively while $V_{\mu\nu} = \partial_\mu V_\nu - \partial_\nu V_\mu$ are the usual bosonic field strengths and the dual field strengths are defined as $\widetilde{V}_{\mu\nu} = \frac{1}{2} \epsilon_{\mu\nu\rho\sigma} V^{\rho \sigma}$.~The effective lagrangian in~\eref{siglag} is composed of the leading terms in a derivative expansion (up to two derivatives) and is useful for parametrizing potentially large new physics effects generated by loops of heavy particles and a convenient framework for assessing the potential sensitivity to the leading operators~\cite{Chen:2014gka} involving photons and $Z$ bosons.

Thus, although~\eref{vertex} is a redundant parameterization of the tensor structure, it is a convenient, yet more general, parametrization for fitting to effective Lagrangian parameters that might be generated in various models at dimension five or less as in~\eref{siglag}.~The parameterization in~\eref{vertex} can of course be mapped, with appropriate translation of the parameters, onto Lagrangians with dimension greater than five or to an underlying dimension six lagrangian in a theory of electroweak symmetry breaking such as in the Standard Model.~We will work explicitly with the vertex in~\eref{vertex} which has been used to calculate the fully differential cross section for $h\rightarrow 4\ell$ and when performing parameter extraction, but other parameterizations can be easily accommodated.

This flexibility of parameterization also allows for other new physics, such as exotic Higgs decays involving exotic fermions or vector bosons~\cite{Falkowski:2014ffa}, to be easily included in the framework.~Furthermore, by using this parameterization, explicit computations of either Standard Model or new physics loop effects which would generate these momentum dependent form factors can easily be included into the framework.~This allows for the ability to in principle extract the parameters from whichever underlying theory is responsible for generating them.~We leave a more detailed investigation of these loop effects to ongoing work~\cite{loopstudy}.

\subsection{Signal and Background Fully\\~~~~~Differential Cross Sections}
\label{subsec:SigandBGCalc}
In the case of signal we have computed analytically the fully differential cross section in the observables described in Sec.\ref{subsec:events} for the process $h \rightarrow ZZ + Z\gamma + \gamma\gamma \rightarrow 4\ell$ using the parameterization in~\eref{vertex}.~We have included all possible interference effects between tensor structures as well as identical final states in the case of $4e/4\mu$.~For the irreducible background we have computed analytically the process $q\bar{q} \rightarrow ZZ + Z\gamma + \gamma\gamma \rightarrow 4\ell$ which includes the s-channel (resonant) $4\ell$ process as well as the t-channel (diboson production) $4\ell$ process and again have included all possible interference effects.~All vector bosons are allowed to be on or off-shell and we do not distinguish between them in what follows.~The details of these calculations can be found in~\cite{Chen:2012jy,Chen:2013ejz,Vega-Morales:2013dda,Chen:2014gka} along with the validation procedures and studies of the distributions as well as the various interference effects.~We have combined these analytic expressions with functions parametrizing the production spectra and implemented them into our analysis framework.

We note that it is important to include all possible Higgs couplings including the $Z\gamma$ and $\gamma\gamma$ contributions in the signal differential cross section since the Higgs appears to be mostly Standard Model-like~\cite{Falkowski:2013dza} and we are primarily searching for small anomalous deviations from the Standard Model prediction.~Thus when attempting to extract specific couplings we must be sure that one small effect is not being mistaken for another.~This is particularly relevant since many of the couplings may be correlated with one another.

Furthermore, it has been shown recently~\cite{Chen:2014gka} that for `true' points near the Standard Model, the greatest sensitivity to the anomalous couplings (non $\mathcal{A}_{1o}^{ZZ}$) is for the $Z\gamma$ and especially $\gamma\gamma$ operators (see~\eref{siglag}).~Including all possible couplings and doing a simultaneous fit ensures that we minimize the possibility of misinterpretation or of introducing a bias when attempting to extract these couplings.~Searching for these small effects is also why it is important to include the interference effects between the identical final state leptons as well as the relevant detector effects and background.

We also comment that in principal there are NLO contributions to the $h\rightarrow 4\ell$ decay processes, but these are expected to be small at $\sim 125$~GeV~\cite{Bredenstein:2006rh,Bredenstein:2006nk} and not relevant until higher precision is obtained once larger data sets are gathered.~Eventually however, these effects should be included and their implementation into our framework is currently ongoing.

\subsection{Combining Production and Decay}
\label{subsec:prodPdec}

To be able to perform a fit for the effective Higgs couplings, we must first construct the fully differential cross section for the observables as a function of the undetermined parameters ($\vec{\mathcal{A}}$).~This differential cross section consists of two components which we assume to be factorized:~the parton level (`decay') differential cross section as discussed in Section \ref{subsec:SigandBGCalc}, and the `production' spectrum.~The full production plus decay fully differential cross section can be expressed as the following, 
\bea
\label{eqn:gen_pdf}
&&P(\vec{p}_{T},Y,\phi, \hat{s},M_{1},M_{2},\vec{\Omega} | \vec{\mathcal{A}}) =\\
&& \frac{dW_{\mathrm{prod}}(\hat{s},\vec{p}_{T},Y,\phi)}{d\hat{s}dYd\vec{p}_{T}d\phi} \times
\frac{d\sigma_{4\ell}(\hat{s},M_1,M_2,\vec{\Omega}|\vec{\mathcal{A}})}{dM_1^2dM_2^2d\vec{\Omega}}, \nonumber
\eea
where, since the Higgs is a spin-0 particle, we can explicitly assume that the decay process can be factorized from the production mechanism.~For the background this explicit factorization does not occur, but still turns out to be an adequate approximation~\cite{Thesis} especially if the $\vec{p}_{T}$ and $Y$ variables are averaged over once the convolution is performed.~The parton level fully differential cross section ($\sigma_{4\ell}$) is treated as being at fixed $\hat{s}$ where one obtains the input $\hat{s}$ value from the production spectrum ($W_{\mathrm{prod}}$).~The production spectrum for the signal and background depend on the parton distribution functions and can not be computed analytically.~For the signal which we assume decays on-shell, the $\hat{s}$ spectrum is taken to be a delta function centered at $m_h^2$, which for a Standard Model Higgs at $125$~GeV is an excellent approximation.~Note however that this assumption can be relaxed in our framework to consider more general $\hat{s}$ spectra as would be found for example in the case of a new heavy scalar with a  large width.

\subsection{Comments Production Spectra}
\label{subsec:prod_specs}
Here we discuss how the $W_{\mathrm{prod}}(\hat{s}, \vec{p}_{T},Y,\phi)$ production spectrum in~\eref{gen_pdf} is obtained.~This function involves higher order effects as well as parton distribution functions and thus can not be computed analytically.~To include them in the total differential cross sections there are various options.~One can in principal generate enough Monte Carlo events to accurately fill the full spectrum in the $(\hat{s}, Y, \vec{p}_T,\phi)$ variables.~As this is computationally intensive we take an approximate approach in which we interpolate analytic functions for the signal and background from one or two dimensional projections generated from the Madgraph~\cite{Alwall:2011uj} and POWHEG~\cite{Melia:2011tj} Monte Carlo generators following a similar procedure as found in~\cite{Anderson:2013fba}.~Having an analytic parameterization for these functions also allows for faster integration when implementing them into the convolution procedure described above.~This procedure of interpolating the one or two dimensional projections neglects correlations between the production variables.~However, since in the signal case there is an explicit factorization between production and decay, the effects of this approximation on parameter extraction in our analysis are small.

In addition, to mitigate these effects further, one can always average over $Y$ and $\vec{p}_T$ as well as fit to ratios of couplings while taking the Higgs mass and overall normalization as input from the total rate (so called `geolocating'~\cite{Gainer:2014hha}) as was done in a recent implementation of our framework into a CMS experimental analysis~\cite{CMS-PAS-HIG-14-014,Khachatryan:2014kca}.~Note however, the overall normalization and Higgs mass can in principle be extracted in our framework, but as this requires extra careful treatment of the production spectra and additional backgrounds we defer a discussion of this to future work.

\subsection{Comments on Additional Backgrounds}
\label{subsec:addbg}

For the background there are also the higher order contributions such as the $gg\rightarrow 4\ell$ and $Z+X$ processes.~These make up the parts of the likelihood which can not currently be included in the convolution integral since a sufficiently accurate analytic parameterization has yet to be obtained.~Thus for these components we must resort to constructing large `look-up' tables via Monte Carlo generation.~Again, the effects of the necessary binning can be mitigated through a linear multi-dimensional interpolation technique~\cite{Thesis}.~Additionally, there will be systematic uncertainties associated with these components.~Fortunately, the $gg\rightarrow 4\ell$ component only makes up $\sim 3-5\%$ relative to $q\bar{q} \rightarrow 4\ell$ around $125$~GeV~\cite{Adam:2008aa}.~The $Z+X$ background on the other hand does make up a sizable contribution of the total background which is comparable to, but smaller than, the largest $q\bar{q}$ component (see Table 2 in~\cite{CMS-PAS-HIG-14-014} or Table 3 in~\cite{Khachatryan:2014kca}).~This component however can in principle be reduced further in the future by requiring more stringent lepton acceptance criteria once more data is collected.~For now the use of the linearly interpolated `look-up' templates and associated systematics is found to be sufficient.~Once the templates are built, including these components in the final likelihood is straightforward as we briefly sketch in~\ssref{sig_back_pdf} and discussed in more detail in~\cite{CMS-PAS-HIG-14-014,Khachatryan:2014kca,Thesis}.


\section{The `Detector Level' \emph{pdf}}
\label{sec:reco_pdf} 

The convolution integral in~\eref{conv} is conceptually straightforward, but in practice is challenging to perform, both for computational and algorithmic reasons.~The key assumption which makes it possible is that the \emph{direction} of lepton momenta are measured with infinite precision which at CMS and ATLAS is a very good approximation.~This allows us, through a change of variables, to reduce the 12-dimensional integral into a more manageable 4-dimensional integral over the four energies of the leptons which are altered by detector resolution effects.~Typically this 4-dimensional integral is done using Monte Carlo techniques~\cite{Artoisenet:2013puc}, thus losing the advantage of having analytic control over the likelihood or assuming that the resolution effects can also be neglecting making the integral trivial.~As discussed in~\sref{overview} we instead perform this integration explicitly using a combination of numerical and analytic methods which allow us to maintain arbitrarily high precision at each step involved.~There are a number of technical details involved in this procedure which are beyond the scope of this `overview' of the framework, but the details can be found in~\cite{Chen:2014hqs,Thesis}.~We instead briefly sketch an overview of the convolution integral and show its validation.

\vspace*{-.33cm}
\subsection{Transforming from CM Basis\\~~~~~~to Lepton Smearing Basis}
\label{subsec:change_var}

Beginning from~\eref{conv} we first discuss the construction of the background detector level \emph{pdf}.~The construction of the signal will be discussed separately as there is a subtle, but important, difference in performing the convolution.~Since there are no undetermined parameters in the background the generator and detector-level (un-normalized) differential cross sections are given simply by $P_B(\vec{X}^G)$ and $P_B(\vec{X}^R)$ respectively and the convolution integral can be written schematically as,
\begin{eqnarray}
\label{eqn:reco_bg}
P_B(\vec{X}^\mathrm{R}) &=& \int P_B(\vec{X}^\mathrm{G} ) T({\vec{X}^{R} | \vec{X}^{G}}) d\vec{X}^\mathrm{G} .
\end{eqnarray}
The set of variables $\vec{X} \equiv (\vec{p}_{T},Y,\phi,\hat{s},M_{1},M_{2},\vec{\Omega})$ exhausts the twelve degrees of freedom (note that $\vec{p}_T$ has 2 components and $\vec{\Omega}$ contains 5 angles) available to the four (massless) final state leptons.~The differential volume element is given by $d\vec{X} = d\hat{s} dM^2_{1}dM^2_{2}d\vec{\Omega} \cdot d\vec{p}_{T}dYd\phi$.

To perform this convolution with the transfer function we must first transform to the basis in which the detector smearing of the lepton momenta is parameterized.~This requires transforming from the basis of the twelve center of mass variables defined in~\ssref{events} to the three momentum basis for the four final state leptons.~In this basis the lepton three momenta $\vec{p}_i$ can be decomposed in terms of the component of the lepton momentum parallel to the direction (${p_i}_{||}$) of motion and the two components perpendicular to the direction of motion ($\vec{p_i}_{\perp}$) (which are zero at generator level).~We then make the assumption that detector smearing will only affect parallel components ${p_i}_{||}$ while the perpendicular components $\vec{p_i}_{\perp}$ are left invariant.~Note that this assumption is equivalent to assuming angular resolution effects due to detector smearing can be neglected,~which is an excellent approximation for the LHC detectors~\cite{CMSperformance2,CMSperformance,CMS-DP-2013-003}.~In the (${p_i}_{||}, \vec{p_i}_{\perp}$) basis only the transfer function associated with ${p_i}_{||}$ is non-trivial while the one associated with the perpendicular components can be represented simply as a delta function for each perpendicular direction, thus allowing for trivial integration over the eight $\vec{p_i}_{\perp}$ variables. 

With these assumptions the integral in~\eref{reco_bg} can then be represented as follows,
\begin{eqnarray}
\label{eqn:reco_pdf3}
P_B(\vec{X}^\mathrm{R} ) 
&=& 
\int P_B(\vec{X}^\mathrm{G}  )
T(\vec{c}~| \vec{P}^{G})  \nonumber\\
&\times& |\mathcal{J}_B| dc_1 dc_3 d{M_1^G}^\mathrm{2} d{M_2^G}^\mathrm{2},
\end{eqnarray}
where we have defined the lepton momenta `smearing factors' $c_i = {p_i}^R_{||}/{p_i}^G_{||}$ and,
\bea
\vec{c} &=& (c_1, c_2, c_3, c_4),~\vec{P}^{G} = (\vec{p_1}^G, \vec{p_2}^G, \vec{p_3}^G, \vec{p_4}^G).
\eea
We have also defined $|\mathcal{J}_B|$ which is the $12\times 12$ Jacobian which parametrizes the (non-linear) transformation that takes us from the center of mass basis to the lepton smearing basis.~The construction of this Jacobian is highly non trivial and requires a combination of analytic and numerical techniques which are beyond the scope of this overview, but the relevant details can be found in~\cite{Chen:2014hqs,Thesis}.

We thus see in~\eref{reco_pdf3} that what started out as a twelve dimensional integral has been reduced to a much more manageable integration over four variables.~The details and validation of this four dimensional integration, which is done using a recursive numerical integration technique~\cite{Bradie:943117} can also be found in~\cite{Chen:2014hqs,Thesis}.

To construct the detector level signal differential cross section (again un-normalized), which is now a function of the effective couplings $\vec{\mathcal{A}}$, we follow the same procedure as for the background starting from,
\bea
\label{eqn:Sconv}
&& P_S(\vec{X}^\mathrm{R} | \vec{\mathcal{A}} ) = 
\int P_S(\vec{X}^\mathrm{G} | \vec{\mathcal{A}} ) 
T({\vec{X}^{R} | \vec{X}^{G}}) d\vec{X}^\mathrm{G}.
\eea
We again use the assumptions which allow us to perform the trivial integration over the eight $\vec{p_i}_{\perp}$ variables, but instead transform to the following integration basis
\begin{eqnarray}
\label{eqn:Sconv2}
P_S(\vec{X}^\mathrm{R} | \vec{\mathcal{A}}) 
&=&  
\int P_S(\vec{X}^\mathrm{G} | \vec{\mathcal{A}}) T(\vec{c}~| \vec{P}^{G}) \nonumber \\ 
&\times& |\mathcal{J}_S|  dc_1 d\hat{s}^G d{M_1^2}^\mathrm{G} d{M_2^2}^\mathrm{G} 
\end{eqnarray}
We now also use the fact that, as mentioned below~\eref{gen_pdf}, the $\hat{s}$ spectrum for the signal is $\propto \delta(\hat{s}^G-m_h^{2})$ (where $m_h$ is the $\emph{generated}$ Higgs mass), enabling us to perform the integration over $d\hat{s}^G$ as well.~Thus, we have for the final signal detector level differential cross section,
\begin{eqnarray}
\label{eqn:Sconv3}
P_S(\vec{X}^\mathrm{R} | \vec{\mathcal{A}}) 
&=&
\int P_S(\vec{X}^\mathrm{G} | \vec{\mathcal{A}}) T(\vec{c}~| \vec{P}^{G})  \\ 
&\times& |\mathcal{J}_S|  dc_1 d{M_1^2}^\mathrm{G} d{M_2^2}^\mathrm{G}\Big|_{\hat{s}^G=m_h^2}, \nonumber
\end{eqnarray}
where again $ |\mathcal{J}_S|$ represent the $12\times 12$ Jacobian (which is different from $|\mathcal{J}_B|$) taking us from the CM basis to the lepton smearing basis.~By using a delta function to model the width of the resonance, there is one less dimension to integrate over as compared to the background case.~While this makes it easier computationally in one respect, an additional complication arises since we have to integrate along a trajectory in which $\hat{s}^G$ is kept constant.~This places an additional constraint when performing the ${M_1^G}^\mathrm{2}$, ${M_2^G}^\mathrm{2}$ integration which further complicates matters and must be properly taken into account.~Explicit details of this integration and its validation along with the derivation of the signal Jacobian $|\mathcal{J}_S|$ in~\eref{Sconv3} are given in~\cite{Chen:2014hqs,Thesis}.

\subsection{Comments on Transfer Function}
\label{subsec:transfer_func}

Detector response effects including effects from selection inefficiency may be parameterized into transfer functions in the following way,
\bea
\label{eqn:tran_fun}
&& T(c_i | \vec{p_i}^G) = \delta(\vec{p_i}_{\perp}^R - \vec{p_i}_{\perp}^G) \nonumber \\
&& \mathcal{S}({p_i}_{||}^R ; {p_i}_{||}^G) \times \epsilon(\vec{p_i}_{\perp}^R,{p_i}_{||}^R),
\eea
The response function $\mathcal{S}$, parameterizes the probability for a lepton with actual momentum $\vec{p}_i^{\mathrm{~G}}$ to be reconstructed with momentum $\vec{p}_i^{\mathrm{~R}}$, while $\delta$ is the Dirac delta function in the perpendicular components, and $\epsilon$ is the selection efficiency.~With typical lepton selection criteria employed by the LHC experiments~\cite{CMSperformance,CMS-DP-2013-003}, it is a good approximation that each lepton is independent.~Thus, the full transfer function for the event may be written as:
\bea
T(\vec{c}~| \vec{P}^{G}) = \prod_{i=1}^4 T(c_i | \vec{p_i}^G).
\eea
We treat $T(\vec{c}~| \vec{P}^{G})$ as a function of $\vec{c}$ which takes the generator level momenta $\vec{P}^{G}$ as input.~The only effect of imperfect momentum measurement on the production spectra is to provide a small smearing of the $\vec{p}_{T}$ spectrum for the four lepton system.~We can mitigate the effects of the smearing by averaging over the production spectra when performing parameter extractions.~Further details on the construction and implementation of the transfer function can be found in~\cite{CMS-PAS-HIG-14-014,Khachatryan:2014kca,Thesis}.

\subsection{Validation of Convolution Integral}
\label{subsec:reco_validation}
As validation of the convolution integral we first show in~\fref{BGconvo_validation1}-\fref{SIGconvo_validation2} projections for signal and background.~We compare in these plots the distributions for a Madgraph sample which has had detector smearing and acceptance effects applied to it versus projections generated from our detector level differential cross sections obtained after the convolution described above.
\begin{figure}
\centering
{\bf~~~~~~~~Detector Level Background Projections}\\
\includegraphics[width=0.235\textwidth]{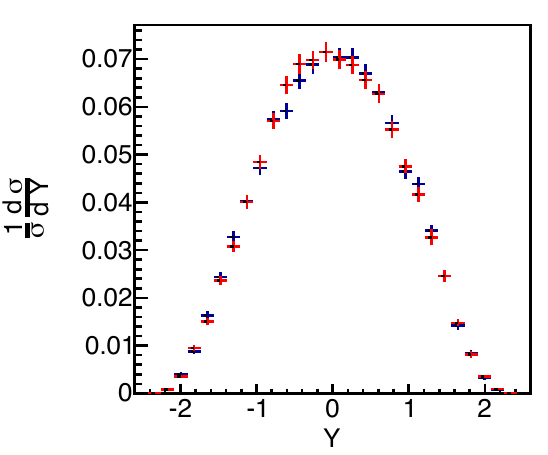}
\includegraphics[width=0.235\textwidth]{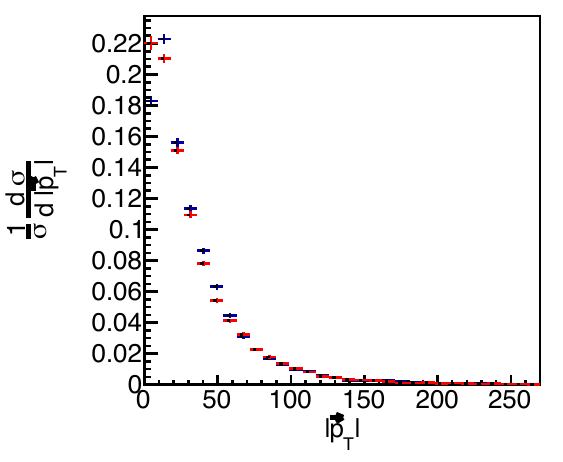}\\
\includegraphics[width=0.235\textwidth]{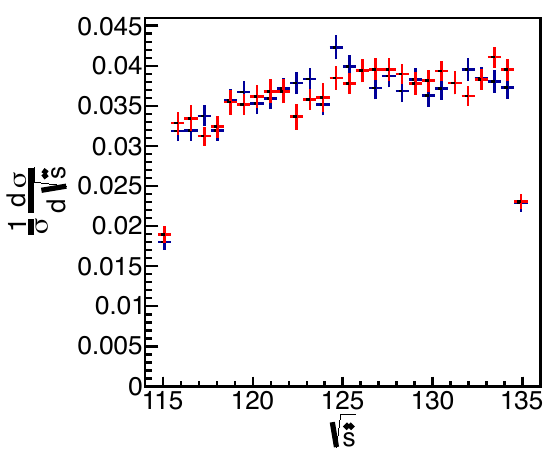}
\includegraphics[width=0.235\textwidth]{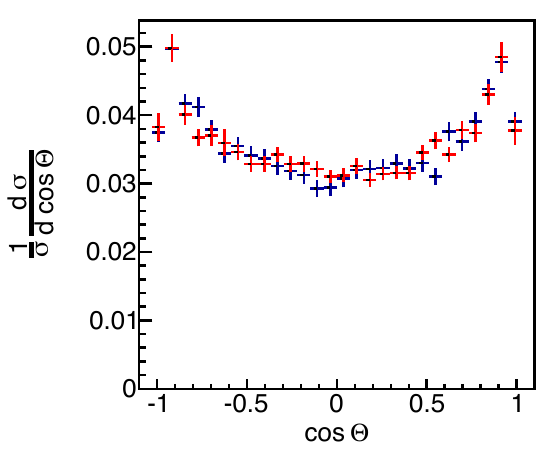}
\caption{Projections of the $Y$, $|\vec{p}_T|$, $\sqrt{\hat{s}} \equiv M_{4\ell}$ and $\cos\Theta$ (see~\ssref{events} for definitions) spectra showing validation of the convolution described in~\eref{reco_bg} for the background.~In \emph{blue} we show the `boosted' Madgraph sample with acceptance cuts and detector smearing applied while in \emph{red} we show projections from our differential cross section after the convolution integration.}
\label{fig:BGconvo_validation1}
\end{figure}
\begin{figure}
\centering
{\bf~~~~~~~~Detector Level Background Projections}\\
\includegraphics[width=0.235\textwidth]{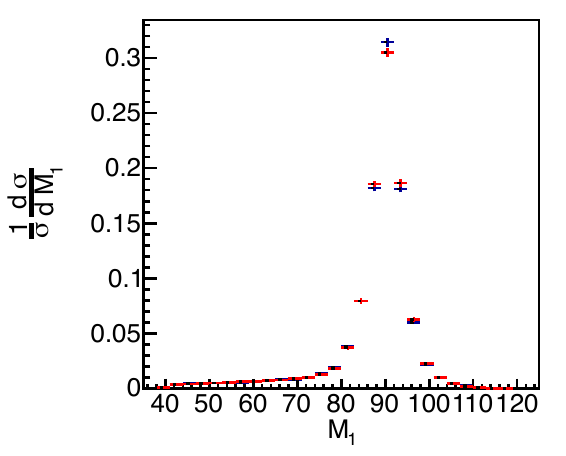}
\includegraphics[width=0.235\textwidth]{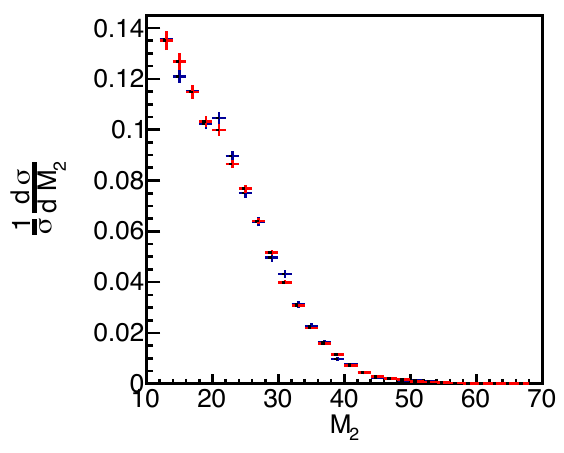}\\
\includegraphics[width=0.235\textwidth]{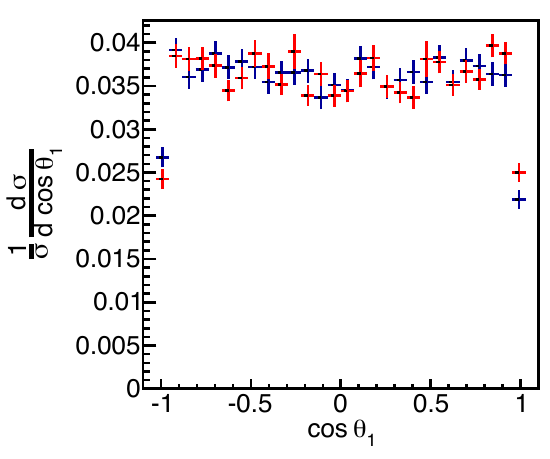}
\includegraphics[width=0.235\textwidth]{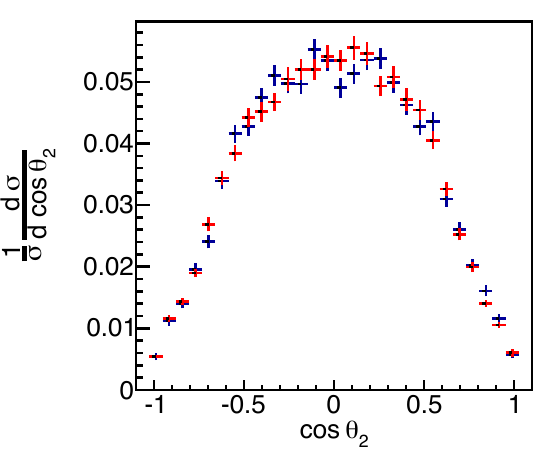}\\
\includegraphics[width=0.235\textwidth]{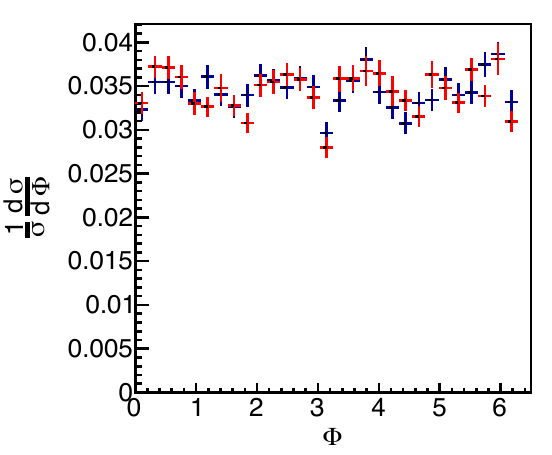}
\includegraphics[width=0.235\textwidth]{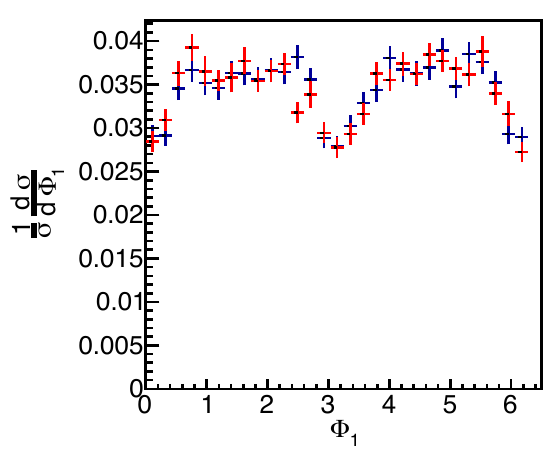}
\caption{Projections of the $M_1$, $M_2$, $\cos\theta_1$, $\cos\theta_2$, $\Phi$ and $\Phi_1$ (see~\ssref{events} for definitions) spectra showing validation of the convolution described in~\eref{reco_bg} for the background.~In \emph{blue} we show the `boosted' Madgraph sample with acceptance cuts and detector smearing applied while in \emph{red} we show projections from our differential cross section after the convolution integration.}
\label{fig:BGconvo_validation2}
\end{figure}
\begin{figure}
\centering
{\bf~~~~~~~~Detector Level Signal Projections}\\
\includegraphics[width=0.235\textwidth]{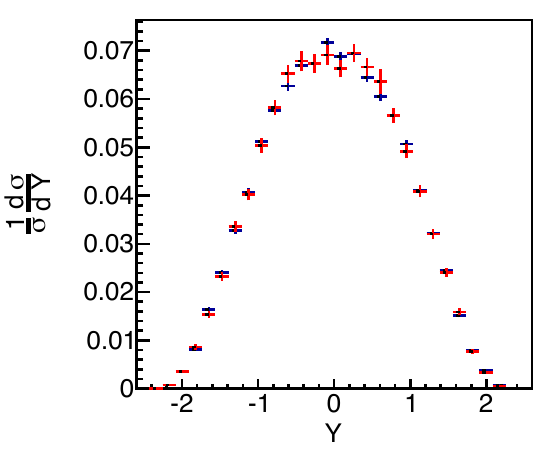}
\includegraphics[width=0.235\textwidth]{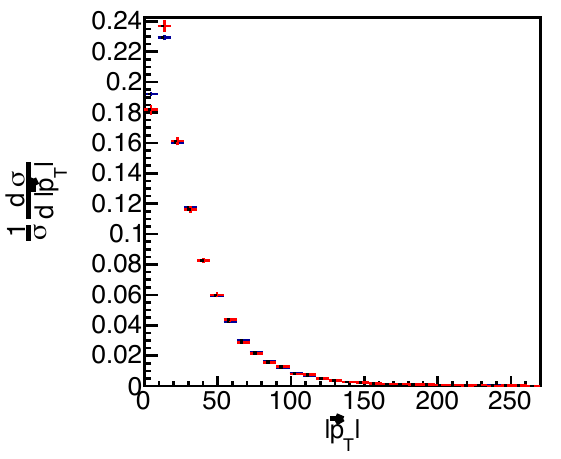}\\
\includegraphics[width=0.235\textwidth]{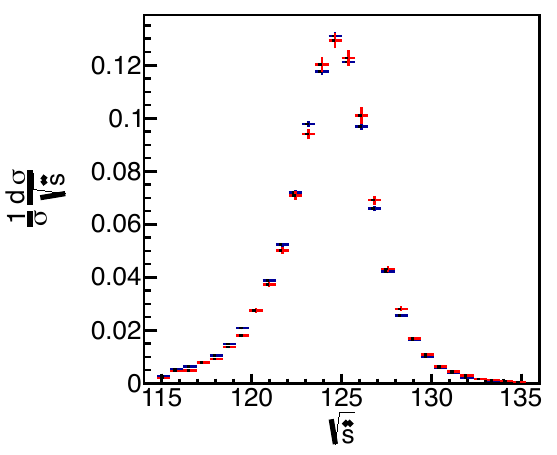}
\includegraphics[width=0.235\textwidth]{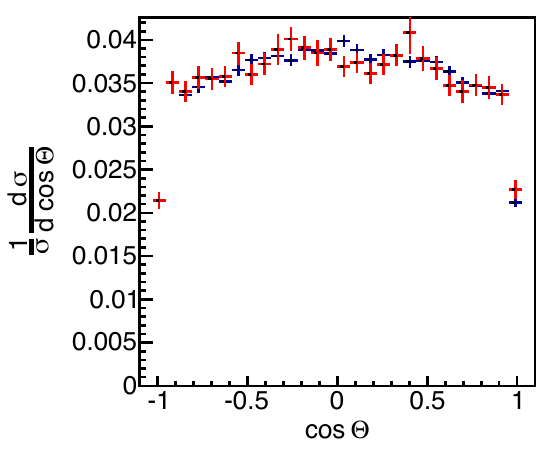}
\caption{Projections of the $Y$, $|\vec{p}_T|$, $\sqrt{\hat{s}} \equiv M_{4\ell}$ and $\cos\Theta$ (see~\ssref{events} for definitions) spectra showing validation of the convolution described in~\eref{Sconv} for the \emph{tree level} SM signal.~In \emph{blue} we show the `boosted' Madgraph sample with acceptance cuts and detector smearing applied while in \emph{red} we show projections from our differential cross section after the convolution integration.}
\label{fig:SIGconvo_validation1}
\end{figure}
\begin{figure}
\centering
{\bf~~~~~~~~Detector Level Signal Projections}\\
\includegraphics[width=0.235\textwidth]{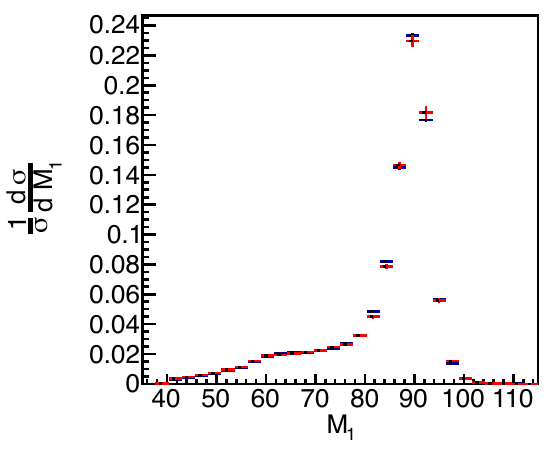}
\includegraphics[width=0.235\textwidth]{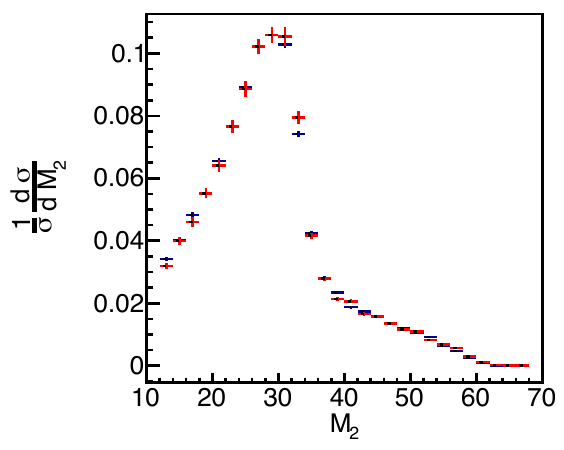}\\
\includegraphics[width=0.235\textwidth]{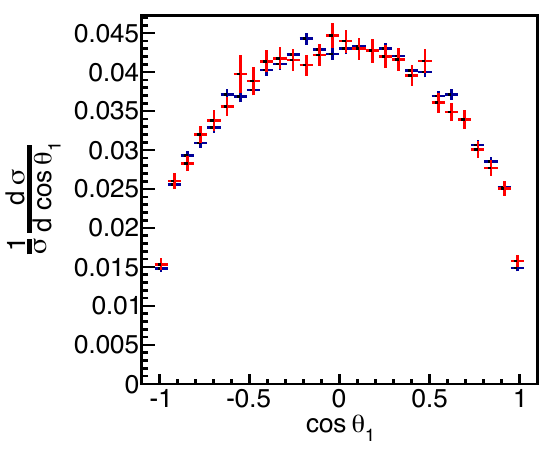}
\includegraphics[width=0.235\textwidth]{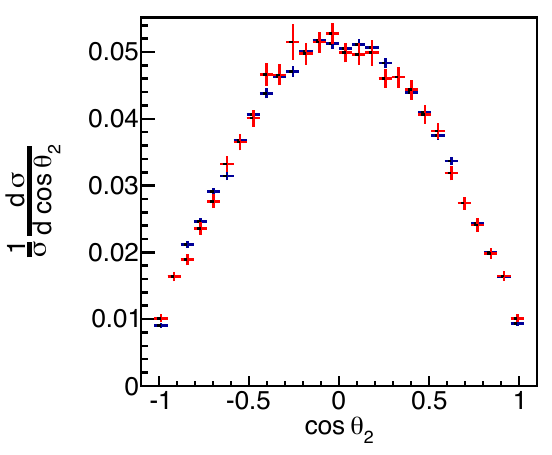}\\
\includegraphics[width=0.235\textwidth]{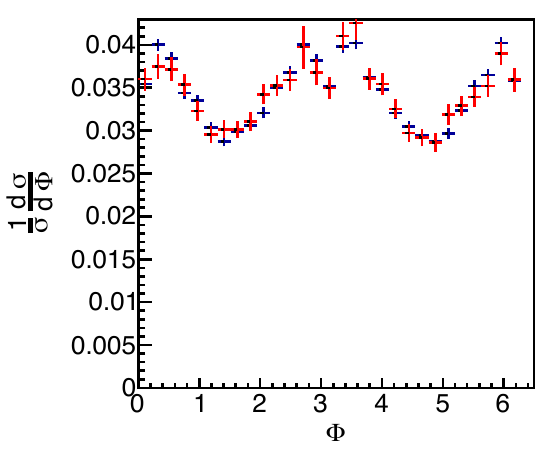}
\includegraphics[width=0.235\textwidth]{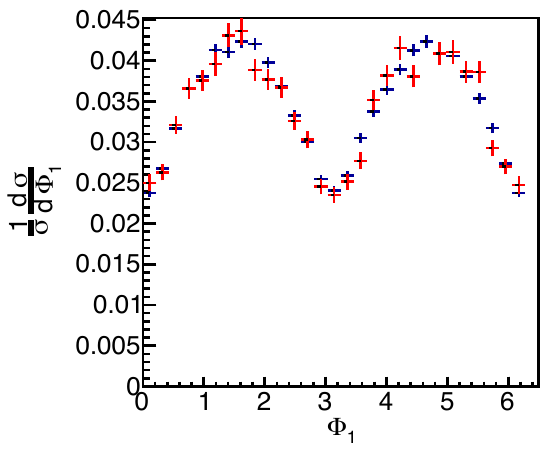}
\caption{Projections of the $M_1$, $M_2$, $\cos\theta_1$, $\cos\theta_2$, $\Phi$ and $\Phi_1$ (see~\ssref{events} for definitions) spectra showing validation of the convolution described in~\eref{Sconv} for the \emph{tree level} SM signal.~In \emph{blue} we show the `boosted' Madgraph sample with acceptance cuts and detector smearing applied while in \emph{red} we show projections from our differential cross section after the convolution integration.}
\label{fig:SIGconvo_validation2}
\end{figure}

We have obtained the signal and background production spectrum for the $(\hat{s}, \vec{p}_T, Y,\phi)$ variables from POWHEG and boosted the Madgraph events and those from our projections accordingly.~We have used the interpolation procedure described in Sec.~\ref{subsec:prod_specs} to build the production spectra for the signal and background differential cross sections and combined them with the analytic expressions for the $h\rightarrow 4\ell$ and $q\bar{q} \rightarrow 4\ell$ processes.~For the signal we show the \emph{tree level} Standard Model point where $\mathcal{A}_{1}^{ZZ} = 2$ and all other couplings are set to zero.~For both signal and background we show only the $2e2\mu$ final state, but results for $4e$ (or $4\mu$) are found in~\cite{CMS-PAS-HIG-14-014,Khachatryan:2014kca,Thesis}.

\begin{figure}
\centering
{\bf~~~~~~Detector Level Background Likelihood}\\
\includegraphics[width=0.42\textwidth]{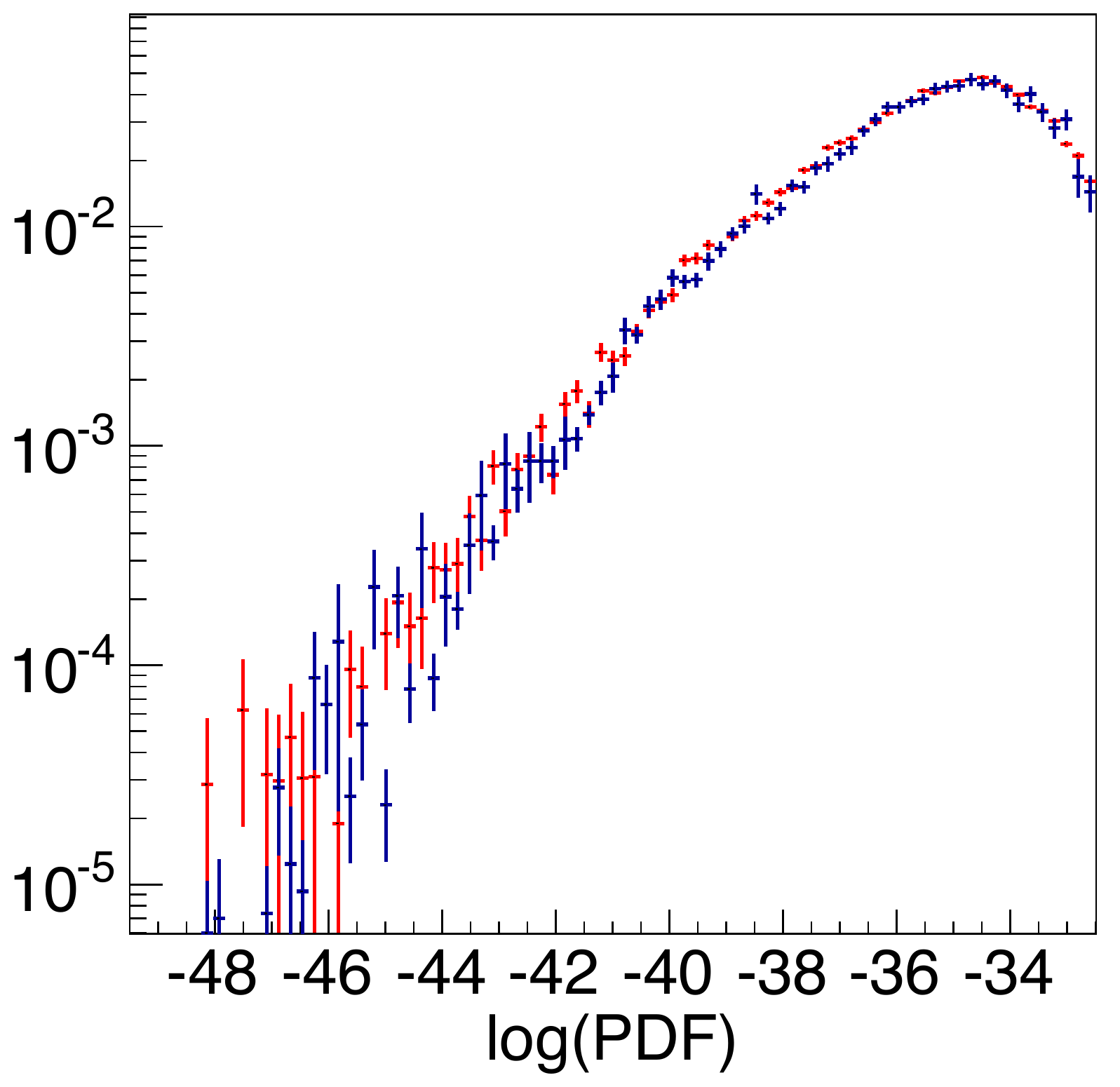}\\
{\bf~~~~~Detector Level Signal Likelihood}\\
\includegraphics[width=0.42\textwidth]{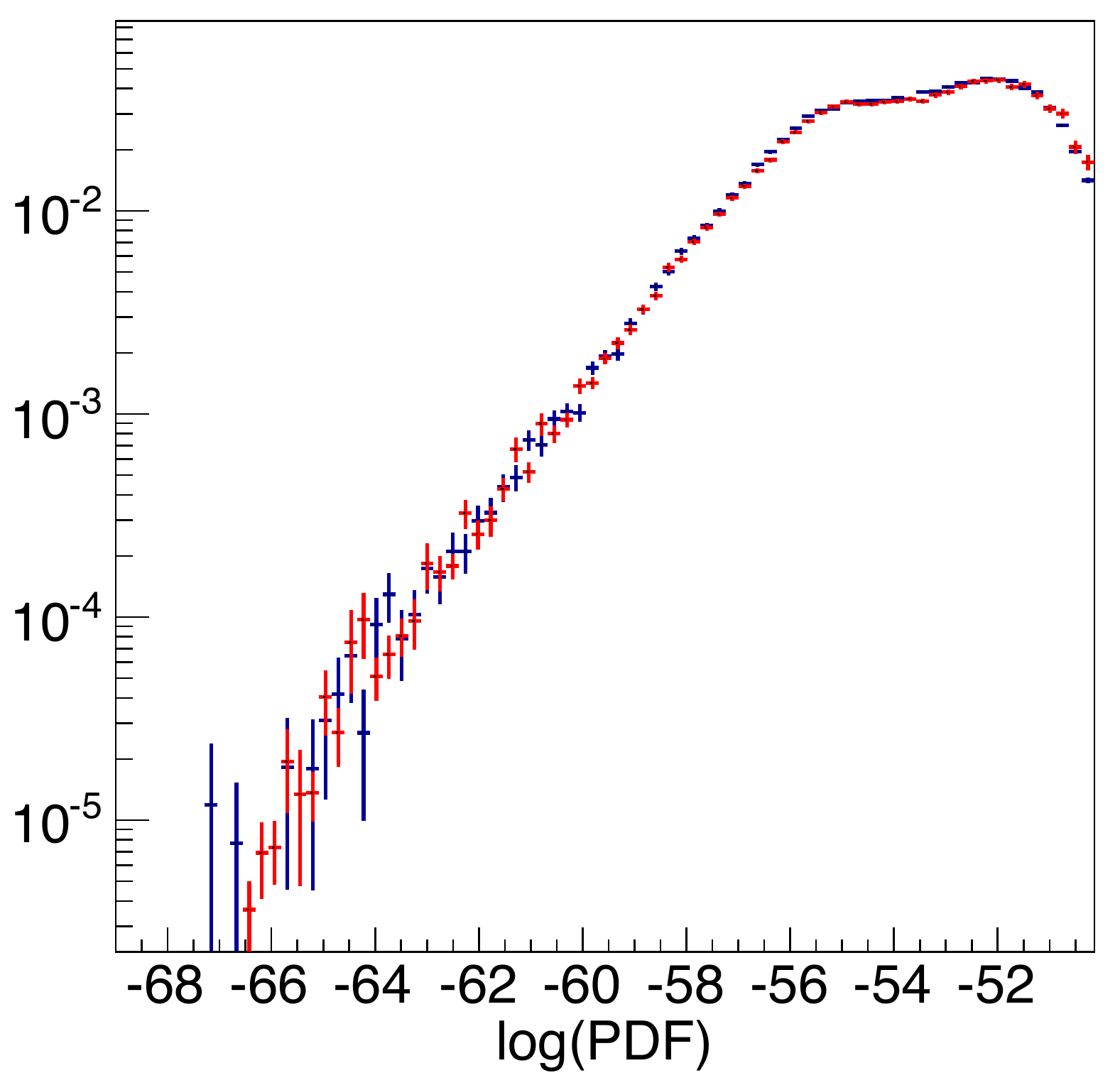}
\caption{Validation of the convolution integrals described in~\eref{reco_bg} and~\eref{Sconv}.~In \emph{blue} we show the `boosted' Madgraph sample with acceptance cuts and detector smearing applied while in \emph{red} we show projections from our differential cross sections after the convolution integration for the \emph{tree level} SM signal and background likelihood.}
\label{fig:convo_validationLH}
\end{figure}

A further validation beyond these projections however is to look at the likelihoods (the differential cross section evaluated for a set of observables) for both the signal and background which contain the full correlations between the different variables.~We show these in Fig.~\ref{fig:convo_validationLH} for a CMS-like phase space and a very large number of events.~To obtain these likelihoods we have evaluated our detector-level differential cross section with the Madgraph sample which has had detector smearing and acceptance effects applied and plotted it on top of the result of evaluating our detector-level differential cross section with events generated from the expression itself.~We find the agreement between the two results to be very good.~Further details are found in the accompanying documents~\cite{Chen:2014hqs,Thesis}.

These plots should not be taken as validation of the complete detector-level differential cross sections which must be validated with full simulation and data.~They are meant only to show the validation of the convolution procedure as well as the construction of the generator-level differential cross sections including the analytic computations.~Complete validations of the full detector level likelihoods including the various production and background effects can be found in~\cite{CMS-PAS-HIG-14-014,Khachatryan:2014kca,Thesis}.

\section{Construction of Likelihoods\\~~~~~~and Parameter Extraction}
\label{sec:likelihoods} 
With the detector level differential cross sections obtained in~\eref{reco_pdf3} and~\eref{Sconv3} in hand we can then go on to construct the full likelihood for a particular dataset.~Before doing so, we must properly normalize the background and signal differential cross sections by performing the full integration over all twelve \emph{reconstructed} $\vec{X}$ variables where from now on we drop the superscript $R$ since we only deal with detector level observables in what follows.~In this section we present a schematic overview of the normalization procedure.~We also at this stage briefly discuss averaging over the production variables $(Y,\vec{p}_T,\phi)$ and the implementation of systematic uncertainties through the use of nuisance parameters in the likelihood functions.~Further details can be found in~\cite{Chen:2014hqs,Thesis}.

\subsection{Normalization of Background and Signal}
\label{subsec:pdf_norm}
One can reduce the effects of production uncertainties by averaging over the detector level production variables ($Y,\vec{p}_T,\phi$).~This is straightforwardly done for the background differential cross sections by the following 4-dimensional integration,
\begin{eqnarray}
\label{eqn:bg_pdf_ptYavg}
P_B(\hat{s},M_1,M_2,\vec{\Omega}) = \int P_B(\vec{X}) dY d\vec{p}_T d\phi.
\end{eqnarray}
An overall volume factor is not shown because for the purpose of likelihood maximization this constant factor is not relevant.~What matters is that the relative normalization between all components in the likelihood is done consistently.~With this differential cross sections in terms of the eight center of mass decay observables we can obtain the overall normalization via a Monte Carlo integration procedure described in~\cite{Chen:2014hqs,Thesis},
\begin{eqnarray}
\label{eqn:bg_norm}
\mathcal{N}_B &=& \int P_B(\hat{s},M_1,M_2,\vec{\Omega}) \nonumber \\
&\times& d\hat{s} d{M^2_1} d{M_2^2} d\vec{\Omega},
\end{eqnarray}
which gives our final normalized background \emph{pdf} as,
\begin{eqnarray}
\label{eqn:bg_norm_pdf}
\mathcal{P}_B(\hat{s},M_1,M_2,\vec{\Omega}) = \mathcal{N}_B^{-1} \times P_B(\hat{s},M_1,M_2,\vec{\Omega}) .
\end{eqnarray}
We have calculated the $q\bar{q} \rightarrow 4\ell$ expression as a sum of the separate individual contributions~\cite{Chen:2012jy,Chen:2013ejz} making it possible to easily perform the integration on each smaller piece to obtain each normalization and then simply sum over them to obtain the overall normalization.

Similarly for the signal we have for the averaging over ($Y, \vec{p}_T,\phi$) variables,
\begin{eqnarray}
\label{eqn:sig_pdf_ptYavg}
P_S(\hat{s},M_1,M_2,\vec{\Omega} | \vec{\mathcal{A}})
= \int P_S(\vec{X} | \vec{\mathcal{A}}) dY d\vec{p}_T.
\end{eqnarray}
To obtain the overall normalization in the signal case we first note that it is a function of the underlying parameters $\vec{\mathcal{A}}$ defined in~\eref{vertex}).~However, from the calculation of the parton level differential cross section presented in~\cite{Chen:2012jy,Chen:2013ejz} or from considering~\eref{vertex} it is clear (assuming constant effective couplings) that $P_S(\hat{s},M_1,M_2,\vec{\Omega} | \vec{\mathcal{A}})$ is a sum over terms each of which is proportional to $\mathcal{A}^{i}_n \mathcal{A}^{j\ast}_{m}$.~Thus we can write,
\begin{eqnarray}
\label{eqn:sig_pdf_sum}
&&P_S(\hat{s},M_1,M_2,\vec{\Omega} | \vec{\mathcal{A}}) = \nonumber \\ 
&& \sum\limits_{ij} \sum\limits_{nm} \mathcal{A}^{i}_n \mathcal{A}^{j\ast}_{m} \times P_S({\hat{s}},M_1,M_2,\vec{\Omega})^{ij}_{nm},
\end{eqnarray}
where $P_S({\hat{s}},M_1,M_2,\vec{\Omega})^{ij}_{nm}$ represents the individual differential cross sections with the couplings factored out.~The separate normalizations for each term can now easily be obtained via,
\begin{eqnarray}
\label{eqn:sig_norm_pieces}
\mathcal{N}^{ij}_{nm} &=& \int P_S(\hat{s},M_1,M_2,\vec{\Omega})^{ij}_{nm} \nonumber \\
&\times& d\hat{s} d{M^2_1} d{M_2^2} d\vec{\Omega} ,
\end{eqnarray}
from which we can now obtain the total overall normalization for the signal \emph{pdf} as,
\begin{eqnarray}
\label{eqn:sig_norm}
\mathcal{N}_S(\vec{\mathcal{A}}) &=& \sum\limits_{ij} \sum\limits_{nm} 
\mathcal{A}^{i}_n \mathcal{A}^{j\ast}_{m} \times \mathcal{N}^{ij}_{nm}. 
\end{eqnarray}
This gives finally for the normalized signal \emph{pdf},
\begin{eqnarray}
\label{eqn:sig_norm_pdf}
&& \mathcal{P}_S({\hat{s}},M_1,M_2,\vec{\Omega} | \vec{\mathcal{A}}) = \nonumber \\
&& \mathcal{N}^{-1}_S(\vec{\mathcal{A}}) \times P_S({\hat{s}},M_1,M_2,\vec{\Omega} | \vec{\mathcal{A}}).
\end{eqnarray}
Since each $\mathcal{N}^{ij}_{nm}$ is computed, one does not need to compute the normalization each time a new hypothesis for $\vec{\mathcal{A}}$ is constructed.~The procedure outlined here also works on more general polynomial functions of the parameters $\vec{\mathcal{A}}$ which one finds after expanding potentially momentum-dependent form factors in powers of momenta.~See~\cite{Chen:2014hqs} for this more general discussion.

Note also that if we take the Higgs mass as a fixed input and only fit for ratios of parameters and not their overall normalization, we do not need the absolute normalization of the differential cross sections.~It thus suffices to have the relative normalization between the different components correct when performing the maximization.~This fact greatly reduces the computational complexity.~Instead of propagating the full normalization and aligning units correctly so that when one integrates over all 8 dimensions unity is obtained, it is sufficient to do a Monte Carlo integration using a fixed sample size in a consistent and sufficiently large range.~The meaning of the log likelihood difference remains unchanged with this construction.~Further details of the normalization procedure for both signal and background are found in~\cite{Chen:2014hqs,Thesis}.

\subsection{Signal Plus Background \emph{pdf}\\~~~~~and Final Likelihood}
\label{subsec:sig_back_pdf}
With~\eref{bg_norm_pdf} and~\eref{sig_norm_pdf} in hand we can now build the signal plus background \emph{pdf} from which the total likelihood will be constructed.~The signal plus background \emph{pdf} can be written as,
\bea
\label{eqn:sigPbg_norm_pdf}
&& \mathcal{P}_{S+B}(\mathcal{O} | F^i_B, \vec{\mathcal{A}}) =  \sum\limits_iF^i_B \times \mathcal{P}^i_{B}(\hat{s},M_1,M_2,\vec{\Omega}) \nonumber \\
&& +~(1-\sum\limits_iF^i_B) \times \mathcal{P}_{S}(\hat{s},M_1,M_2,\vec{\Omega} | \vec{\mathcal{A}}). 
\eea
where $\mathcal{O} \equiv (\hat{s},M_1,M_2,\vec{\Omega})$ is our final set of observables to be used in the construction of the likelihood and $F_B^i$ is the background fraction for a particular component, each of which must also be extracted.

The sum over background components is given by,
\bea
\label{eqn:fbg_pdf_sum}
\sum\limits_iF^i_B &\times& \mathcal{P}^i_{B}(\mathcal{O})
= F_B^{q\bar{q}} P^{q\bar{q}}_B(\mathcal{O})   \nonumber \\
&+& F_B^{gg} P^{gg}_B(\mathcal{O}) 
+ F_B^{Z+X} P^{Z+X}_B(\mathcal{O}), 
\eea
where $P^{q\bar{q}}_B(\mathcal{O})$ is the dominant $q\bar{q}\to4\ell$ component and is obtained via the convolution integral in~\eref{conv}.~The sub-dominant $gg\to4\ell$ and $Z+X$ components, given by $P^{gg}_B(\mathcal{O})$ and $P^{Z+X}_B(\mathcal{O})$ respectively, must be obtained via the linearly interpolated `look-up' tables from large Monte Carlo samples as discussed in~\ssref{addbg}.

We can now write the likelihood of obtaining a particular dataset containing $N$ events as,
\begin{eqnarray}
\label{eqn:likelihood}
&& L(F_B^i, \vec{\mathcal{A}}) = \prod_{\mathcal{O}}^N \mathcal{P}_{S+B}(\mathcal{O} | F_B^i, \vec{\mathcal{A}}).
\end{eqnarray}
This likelihood can also be combined with an appropriate poisson weighting factor to account for the probability of observing a given number of events~\cite{CMS-PAS-HIG-14-014,Khachatryan:2014kca,Thesis}.~In the case of multiple final states (for example $4e$, $4\mu$ and $2e2\mu$), we build the likelihood function and implement the appropriate systematic uncertainties for each one separately.~We now briefly discuss the implementation of the systematic uncertainties.

\subsection{Including Systematic Uncertainties}
\label{subsec:nuisance}
Systematic uncertainties must be accounted for given our  imperfect knowledge of various aspects of the analysis procedure.~The lepton momentum resolution, the size of the backgrounds, and the exact production spectra are some important examples.~For each of these systematic uncertainties we can associate an undetermined parameter which parametrizes our ignorance of the corresponding effect.~Since we are not directly interested in these parameters, but only use them to estimate our systematic uncertainties, they are deemed nuisance parameters and are subsequently profiled over~\cite{CMS-PAS-HIG-14-014,Khachatryan:2014kca}.

This is done by generating alternative \emph{pdfs} using different values for the nuisance parameter of interest.~To give one important example, we generate  \emph{pdfs} with narrower or wider lepton response functions to parameterize our knowledge of the lepton momentum resolution.~If we define the nominal \emph{pdf} to be $\mathcal{P}_0(\mathcal{O})$ and the alternative as $\mathcal{P}_1(\mathcal{O})$, one can parameterize the dependence of the likelihood on a nuisance parameter $n$ by interpolating between the nominal and the alternative \emph{pdfs} as follows:
\begin{eqnarray}
\mathcal{P}(\mathcal{O}|n) &=& (1 - n)\, \mathcal{P}_0(\mathcal{O})
+ n \, P_1(\mathcal{O}) \nonumber\\
&=& P_0(\mathcal{O}) + n \left[P_1(\mathcal{O}) - P_0(\mathcal{O})\right].
\end{eqnarray}
It is instructive to observe that, for all values of $n$, the normalization of the total \emph{pdf} stays the same.~Given the asymmetric nature of many systematic uncertainties, it is more appropriate to generate many ``check-points'' along the axis of $n$ and to do piece-wise interpolation without the need of worrying about the normalization.~Non-central values of $n$ are \textit{a priori} disfavored, therefore one can impose a prior on top of the interpolated likelihood:
\begin{equation}
\mathcal{P}(\mathcal{O} | n) = \mathcal{P}(\mathcal{O} | n) \mathcal{G}(n),
\end{equation}
where $\mathcal{G}(n)$ is typically a Gaussian centered at the central value of $n$.~In the case of multiple systematic uncertainties, one can replace $n$ by a vector of nuisance parameters $\vec{n}$, and the prior $\mathcal{G}(n)$ by $\mathcal{G}(\vec{n})$.~In general $\mathcal{G}(\vec{n})$ is a multivariate Gaussian-like function with primary axes which are some combination of different nuisance parameter directions.~However one can carefully define the nuisance parameters such that correlations between them are negligible.~In this limit $\mathcal{G}(\vec{n})$ can be written as the product of many Gaussian-like functions.~This procedure for including systematic uncertainties has been implemented in a recent CMS analysis utilizing our framework~\cite{CMS-PAS-HIG-14-014,Khachatryan:2014kca} and further details can be found in~\cite{Thesis}.

\subsection{Comments on Parameter Extraction}
\label{subsec:param_ext}

As discussed in~\cite{Gao:2010qx,Bolognesi:2012mm,Anderson:2013fba} the advantage of analytic approaches is that the likelihood can be maximized for a large set of parameters in the most optimal way without losing information.~Our framework allows for the `analytic' nature of these approaches to be maintained at detector level giving us the ability to perform fast and accurate multi-parameter fits for lagrangian parameters directly from the data.~This is possible once the convolution in~\eref{conv} is performed and after normalization of the signal and background $\emph{pdfs}$ allowing us to obtain the full detector level likelihood $L(\vec{\mathcal{A}})$ for a particular dataset.~With the likelihood in hand a maximization procedure to find the \emph{global} maximum can be performed to obtain the value of the parameters for which the likelihood is maximized.~For this task we have incorporated the well established MINUIT~\cite{James:1994vla} function minimization/maximization code into our framework.~We find excellent rates of convergence and a high degree of stability in locating the global maximum of the likelihood as well as accurate extraction of the parameters as demonstrated in~\cite{CMS-PAS-HIG-14-014,Khachatryan:2014kca,Chen:2014hqs,Thesis} where more details can be found.

One important feature of the procedure is that the computationally intensive component of evaluating the likelihood only needs to be done for the events in the final dataset used in the fit for a given experiment.~Therefore the computationally expensive pieces can be calculated on the computing grid \emph{prior to the analysis of the data}, and the fit for parameter extraction itself is then completed within a few seconds.~This allows for a great deal of flexibility, including testing alternative parameterizations, when fitting the undetermined parameters.~Many examples of the types of parameter extractions which can be done within our framework, both at generator and at detector level, can be found in~\cite{Chen:2013ejz,Chen:2014gka,CMS-PAS-HIG-14-014,Khachatryan:2014kca,Thesis}. 

\section{Summary and Conclusions}
\label{sec:conc} 

In this study we build upon an earlier study~\cite{Chen:2013ejz} to construct a comprehensive analysis framework aimed at extracting as much information as possible from the Higgs golden channel.~Our framework is based on a maximum likelihood method constructed from analytic expressions of the fully differential cross sections for the $h \rightarrow 4\ell$ decay as well as the dominant irreducible $q\bar{q} \rightarrow 4\ell$ background which were computed in~\cite{Chen:2012jy,Chen:2013ejz}.~As our main result, we have constructed the full 12-dimensional detector level likelihood utilizing all observables available in the golden channel.~This allows us to perform parameter extraction of the various possible Higgs couplings, including general CP odd/even admixtures and any possible phases.

The detector-level likelihood is obtained by the explicit convolution of a transfer function, encapsulating the relevant detector effects, with the generator-level probability density formed out of the signal and background differential cross sections.~After performing this 12-dimensional convolution integral and its normalization we obtain a probability density function from which we construct an un-binned detector-level likelihood which is a continuous function of the effective couplings.

In summary we have given broad overview of a framework optimized for extracting Higgs couplings in the golden channel.~We have sketched how the convolution is performed and shown various validations as well as discussed the principles and theoretical basis of the framework.~Many of the technical details as well as results using our framework can be found in~\cite{Chen:2012jy,Chen:2013ejz,Chen:2014gka,CMS-PAS-HIG-14-014,Chen:2014hqs,Thesis}.~This framework has already proved useful in a recent CMS analysis~\cite{CMS-PAS-HIG-14-014,Khachatryan:2014kca} and can be used in the future in a variety of ways to study Higgs couplings in the golden channel using data obtained at the LHC and other future colliders. 

\vskip 0.2 cm

\noindent
{\bf Acknowledgments:} The authors are grateful to Artur Apresyan, Michalis Bachtis, Adam Falkowski, Patrick Fox, Andrei Gritsan, Roni Harnik, Alex Mott, Pavel Nadolsky, Daniel Stolarski, Reisaburo Tanaka, Nhan Tran, Roberto Vega, and Felix Yu for helpful discussions as well as the ATLAS and CMS collaborations for their encouragement and interest in this work.~R.V.M.~is supported by the Fermilab Graduate Student Fellowship in Theoretical Physics and the ERC Advanced Grant Higgs@LHC.~Fermilab is operated by Fermi Research Alliance, LLC, under Contract No.~DE-AC02-07CH11359 with the United States Department of Energy.~Y.C.~is supported by the Weston Havens Foundation and DOE grant DE-FG02-92-ER-40701.~This work is also sponsored in part by the DOE grant No.~DE-FG02-91ER40684.~R.V.M is also grateful to the CalTech physics department for their hospitality during which much of this work was done.~This work used the Extreme Science and Engineering Discovery Environment (XSEDE), which is supported by National Science Foundation grant number OCI-1053575.


\bibliographystyle{apsrev}
\bibliography{GoldenChannelBib}

\begin{thebibliography}{63}
\expandafter\ifx\csname natexlab\endcsname\relax\def\natexlab#1{#1}\fi
\expandafter\ifx\csname bibnamefont\endcsname\relax
  \def\bibnamefont#1{#1}\fi
\expandafter\ifx\csname bibfnamefont\endcsname\relax
  \def\bibfnamefont#1{#1}\fi
\expandafter\ifx\csname citenamefont\endcsname\relax
  \def\citenamefont#1{#1}\fi
\expandafter\ifx\csname url\endcsname\relax
  \def\url#1{\texttt{#1}}\fi
\expandafter\ifx\csname urlprefix\endcsname\relax\def\urlprefix{URL }\fi
\providecommand{\bibinfo}[2]{#2}
\providecommand{\eprint}[2][]{\url{#2}}

\bibitem[{\citenamefont{Aad et~al.}(2012)}]{:2012gk}
\bibinfo{author}{\bibfnamefont{G.}~\bibnamefont{Aad}} \bibnamefont{et~al.}
  (\bibinfo{collaboration}{ATLAS Collaboration}), \bibinfo{journal}{Phys.Lett.}
  \textbf{\bibinfo{volume}{B716}}, \bibinfo{pages}{1} (\bibinfo{year}{2012}),
  \eprint{1207.7214}.

\bibitem[{\citenamefont{Chatrchyan et~al.}(2012{\natexlab{a}})}]{:2012gu}
\bibinfo{author}{\bibfnamefont{S.}~\bibnamefont{Chatrchyan}}
  \bibnamefont{et~al.} (\bibinfo{collaboration}{CMS Collaboration}),
  \bibinfo{journal}{Phys.Lett.} \textbf{\bibinfo{volume}{B716}},
  \bibinfo{pages}{30} (\bibinfo{year}{2012}{\natexlab{a}}), \eprint{1207.7235}.

\bibitem[{\citenamefont{Nelson}(1988)}]{Nelson:1986ki}
\bibinfo{author}{\bibfnamefont{C.~A.} \bibnamefont{Nelson}},
  \bibinfo{journal}{Phys.Rev.} \textbf{\bibinfo{volume}{D37}},
  \bibinfo{pages}{1220} (\bibinfo{year}{1988}).

\bibitem[{\citenamefont{Soni and Xu}(1993)}]{Soni:1993jc}
\bibinfo{author}{\bibfnamefont{A.}~\bibnamefont{Soni}} \bibnamefont{and}
  \bibinfo{author}{\bibfnamefont{R.}~\bibnamefont{Xu}},
  \bibinfo{journal}{Phys.Rev.} \textbf{\bibinfo{volume}{D48}},
  \bibinfo{pages}{5259} (\bibinfo{year}{1993}), \eprint{hep-ph/9301225}.

\bibitem[{\citenamefont{Chang et~al.}(1993)\citenamefont{Chang, Keung, and
  Phillips}}]{Chang:1993jy}
\bibinfo{author}{\bibfnamefont{D.}~\bibnamefont{Chang}},
  \bibinfo{author}{\bibfnamefont{W.-Y.} \bibnamefont{Keung}}, \bibnamefont{and}
  \bibinfo{author}{\bibfnamefont{I.}~\bibnamefont{Phillips}},
  \bibinfo{journal}{Phys.Rev.} \textbf{\bibinfo{volume}{D48}},
  \bibinfo{pages}{3225} (\bibinfo{year}{1993}), \eprint{hep-ph/9303226}.

\bibitem[{\citenamefont{Barger et~al.}(1994)\citenamefont{Barger, Cheung,
  Djouadi, Kniehl, and Zerwas}}]{Barger:1993wt}
\bibinfo{author}{\bibfnamefont{V.~D.} \bibnamefont{Barger}},
  \bibinfo{author}{\bibfnamefont{K.-m.} \bibnamefont{Cheung}},
  \bibinfo{author}{\bibfnamefont{A.}~\bibnamefont{Djouadi}},
  \bibinfo{author}{\bibfnamefont{B.~A.} \bibnamefont{Kniehl}},
  \bibnamefont{and} \bibinfo{author}{\bibfnamefont{P.}~\bibnamefont{Zerwas}},
  \bibinfo{journal}{Phys.Rev.} \textbf{\bibinfo{volume}{D49}},
  \bibinfo{pages}{79} (\bibinfo{year}{1994}), \eprint{hep-ph/9306270}.

\bibitem[{\citenamefont{Arens and Sehgal}(1995)}]{Arens:1994wd}
\bibinfo{author}{\bibfnamefont{T.}~\bibnamefont{Arens}} \bibnamefont{and}
  \bibinfo{author}{\bibfnamefont{L.}~\bibnamefont{Sehgal}},
  \bibinfo{journal}{Z.Phys.} \textbf{\bibinfo{volume}{C66}},
  \bibinfo{pages}{89} (\bibinfo{year}{1995}), \eprint{hep-ph/9409396}.

\bibitem[{\citenamefont{Choi et~al.}(2003)\citenamefont{Choi, Miller,
  Muhlleitner, and Zerwas}}]{Choi:2002jk}
\bibinfo{author}{\bibfnamefont{S.}~\bibnamefont{Choi}},
  \bibinfo{author}{\bibfnamefont{.}~\bibnamefont{Miller}, \bibfnamefont{D.J.}},
  \bibinfo{author}{\bibfnamefont{M.}~\bibnamefont{Muhlleitner}},
  \bibnamefont{and} \bibinfo{author}{\bibfnamefont{P.}~\bibnamefont{Zerwas}},
  \bibinfo{journal}{Phys.Lett.} \textbf{\bibinfo{volume}{B553}},
  \bibinfo{pages}{61} (\bibinfo{year}{2003}), \eprint{hep-ph/0210077}.

\bibitem[{\citenamefont{Buszello et~al.}(2004)\citenamefont{Buszello, Fleck,
  Marquard, and van~der Bij}}]{Buszello:2002uu}
\bibinfo{author}{\bibfnamefont{C.}~\bibnamefont{Buszello}},
  \bibinfo{author}{\bibfnamefont{I.}~\bibnamefont{Fleck}},
  \bibinfo{author}{\bibfnamefont{P.}~\bibnamefont{Marquard}}, \bibnamefont{and}
  \bibinfo{author}{\bibfnamefont{J.}~\bibnamefont{van~der Bij}},
  \bibinfo{journal}{Eur.Phys.J.} \textbf{\bibinfo{volume}{C32}},
  \bibinfo{pages}{209} (\bibinfo{year}{2004}), \eprint{hep-ph/0212396}.

\bibitem[{\citenamefont{Godbole et~al.}(2007)\citenamefont{Godbole, Miller, and
  Muhlleitner}}]{Godbole:2007cn}
\bibinfo{author}{\bibfnamefont{R.~M.} \bibnamefont{Godbole}},
  \bibinfo{author}{\bibfnamefont{.}~\bibnamefont{Miller}, \bibfnamefont{D.J.}},
  \bibnamefont{and} \bibinfo{author}{\bibfnamefont{M.~M.}
  \bibnamefont{Muhlleitner}}, \bibinfo{journal}{JHEP}
  \textbf{\bibinfo{volume}{0712}}, \bibinfo{pages}{031} (\bibinfo{year}{2007}),
  \eprint{0708.0458}.

\bibitem[{\citenamefont{Kovalchuk}(2008)}]{Kovalchuk:2008zz}
\bibinfo{author}{\bibfnamefont{V.}~\bibnamefont{Kovalchuk}},
  \bibinfo{journal}{J.Exp.Theor.Phys.} \textbf{\bibinfo{volume}{107}},
  \bibinfo{pages}{774} (\bibinfo{year}{2008}).

\bibitem[{\citenamefont{Cao et~al.}(2010)\citenamefont{Cao, Jackson, Keung,
  Low, and Shu}}]{Cao:2009ah}
\bibinfo{author}{\bibfnamefont{Q.-H.} \bibnamefont{Cao}},
  \bibinfo{author}{\bibfnamefont{C.}~\bibnamefont{Jackson}},
  \bibinfo{author}{\bibfnamefont{W.-Y.} \bibnamefont{Keung}},
  \bibinfo{author}{\bibfnamefont{I.}~\bibnamefont{Low}}, \bibnamefont{and}
  \bibinfo{author}{\bibfnamefont{J.}~\bibnamefont{Shu}},
  \bibinfo{journal}{Phys.Rev.} \textbf{\bibinfo{volume}{D81}},
  \bibinfo{pages}{015010} (\bibinfo{year}{2010}), \eprint{0911.3398}.

\bibitem[{\citenamefont{Gao et~al.}(2010)\citenamefont{Gao, Gritsan, Guo,
  Melnikov, Schulze et~al.}}]{Gao:2010qx}
\bibinfo{author}{\bibfnamefont{Y.}~\bibnamefont{Gao}},
  \bibinfo{author}{\bibfnamefont{A.~V.} \bibnamefont{Gritsan}},
  \bibinfo{author}{\bibfnamefont{Z.}~\bibnamefont{Guo}},
  \bibinfo{author}{\bibfnamefont{K.}~\bibnamefont{Melnikov}},
  \bibinfo{author}{\bibfnamefont{M.}~\bibnamefont{Schulze}},
  \bibnamefont{et~al.}, \bibinfo{journal}{Phys.Rev.}
  \textbf{\bibinfo{volume}{D81}}, \bibinfo{pages}{075022}
  (\bibinfo{year}{2010}), \eprint{1001.3396}.

\bibitem[{\citenamefont{De~Rujula et~al.}(2010)\citenamefont{De~Rujula, Lykken,
  Pierini, Rogan, and Spiropulu}}]{DeRujula:2010ys}
\bibinfo{author}{\bibfnamefont{A.}~\bibnamefont{De~Rujula}},
  \bibinfo{author}{\bibfnamefont{J.}~\bibnamefont{Lykken}},
  \bibinfo{author}{\bibfnamefont{M.}~\bibnamefont{Pierini}},
  \bibinfo{author}{\bibfnamefont{C.}~\bibnamefont{Rogan}}, \bibnamefont{and}
  \bibinfo{author}{\bibfnamefont{M.}~\bibnamefont{Spiropulu}},
  \bibinfo{journal}{Phys.Rev.} \textbf{\bibinfo{volume}{D82}},
  \bibinfo{pages}{013003} (\bibinfo{year}{2010}), \eprint{1001.5300}.

\bibitem[{\citenamefont{Gainer et~al.}(2011)\citenamefont{Gainer, Kumar, Low,
  and Vega-Morales}}]{Gainer:2011xz}
\bibinfo{author}{\bibfnamefont{J.~S.} \bibnamefont{Gainer}},
  \bibinfo{author}{\bibfnamefont{K.}~\bibnamefont{Kumar}},
  \bibinfo{author}{\bibfnamefont{I.}~\bibnamefont{Low}}, \bibnamefont{and}
  \bibinfo{author}{\bibfnamefont{R.}~\bibnamefont{Vega-Morales}},
  \bibinfo{journal}{JHEP} \textbf{\bibinfo{volume}{1111}}, \bibinfo{pages}{027}
  (\bibinfo{year}{2011}), \eprint{1108.2274}.

\bibitem[{\citenamefont{Coleppa et~al.}(2012)\citenamefont{Coleppa, Kumar, and
  Logan}}]{Coleppa:2012eh}
\bibinfo{author}{\bibfnamefont{B.}~\bibnamefont{Coleppa}},
  \bibinfo{author}{\bibfnamefont{K.}~\bibnamefont{Kumar}}, \bibnamefont{and}
  \bibinfo{author}{\bibfnamefont{H.~E.} \bibnamefont{Logan}}
  (\bibinfo{year}{2012}), \eprint{1208.2692}.

\bibitem[{\citenamefont{Bolognesi et~al.}(2012)\citenamefont{Bolognesi, Gao,
  Gritsan, Melnikov, Schulze et~al.}}]{Bolognesi:2012mm}
\bibinfo{author}{\bibfnamefont{S.}~\bibnamefont{Bolognesi}},
  \bibinfo{author}{\bibfnamefont{Y.}~\bibnamefont{Gao}},
  \bibinfo{author}{\bibfnamefont{A.~V.} \bibnamefont{Gritsan}},
  \bibinfo{author}{\bibfnamefont{K.}~\bibnamefont{Melnikov}},
  \bibinfo{author}{\bibfnamefont{M.}~\bibnamefont{Schulze}},
  \bibnamefont{et~al.} (\bibinfo{year}{2012}), \eprint{1208.4018}.

\bibitem[{\citenamefont{Stolarski and Vega-Morales}(2012)}]{Stolarski:2012ps}
\bibinfo{author}{\bibfnamefont{D.}~\bibnamefont{Stolarski}} \bibnamefont{and}
  \bibinfo{author}{\bibfnamefont{R.}~\bibnamefont{Vega-Morales}},
  \bibinfo{journal}{Phys.Rev.} \textbf{\bibinfo{volume}{D86}},
  \bibinfo{pages}{117504} (\bibinfo{year}{2012}), \eprint{1208.4840}.

\bibitem[{\citenamefont{Chen et~al.}(2013{\natexlab{a}})\citenamefont{Chen,
  Tran, and Vega-Morales}}]{Chen:2012jy}
\bibinfo{author}{\bibfnamefont{Y.}~\bibnamefont{Chen}},
  \bibinfo{author}{\bibfnamefont{N.}~\bibnamefont{Tran}}, \bibnamefont{and}
  \bibinfo{author}{\bibfnamefont{R.}~\bibnamefont{Vega-Morales}},
  \bibinfo{journal}{JHEP} \textbf{\bibinfo{volume}{1301}}, \bibinfo{pages}{182}
  (\bibinfo{year}{2013}{\natexlab{a}}), \eprint{1211.1959}.

\bibitem[{\citenamefont{Boughezal et~al.}(2012)\citenamefont{Boughezal,
  LeCompte, and Petriello}}]{Boughezal:2012tz}
\bibinfo{author}{\bibfnamefont{R.}~\bibnamefont{Boughezal}},
  \bibinfo{author}{\bibfnamefont{T.~J.} \bibnamefont{LeCompte}},
  \bibnamefont{and} \bibinfo{author}{\bibfnamefont{F.}~\bibnamefont{Petriello}}
  (\bibinfo{year}{2012}), \eprint{1208.4311}.

\bibitem[{\citenamefont{Belyaev et~al.}(2012)\citenamefont{Belyaev,
  Christensen, and Pukhov}}]{Belyaev:2012qa}
\bibinfo{author}{\bibfnamefont{A.}~\bibnamefont{Belyaev}},
  \bibinfo{author}{\bibfnamefont{N.~D.} \bibnamefont{Christensen}},
  \bibnamefont{and} \bibinfo{author}{\bibfnamefont{A.}~\bibnamefont{Pukhov}}
  (\bibinfo{year}{2012}), \eprint{1207.6082}.

\bibitem[{\citenamefont{Avery et~al.}(2012)\citenamefont{Avery, Bourilkov,
  Chen, Cheng, Drozdetskiy et~al.}}]{Avery:2012um}
\bibinfo{author}{\bibfnamefont{P.}~\bibnamefont{Avery}},
  \bibinfo{author}{\bibfnamefont{D.}~\bibnamefont{Bourilkov}},
  \bibinfo{author}{\bibfnamefont{M.}~\bibnamefont{Chen}},
  \bibinfo{author}{\bibfnamefont{T.}~\bibnamefont{Cheng}},
  \bibinfo{author}{\bibfnamefont{A.}~\bibnamefont{Drozdetskiy}},
  \bibnamefont{et~al.} (\bibinfo{year}{2012}), \eprint{1210.0896}.

\bibitem[{\citenamefont{Campbell
  et~al.}(2012{\natexlab{a}})\citenamefont{Campbell, Giele, and
  Williams}}]{Campbell:2012ct}
\bibinfo{author}{\bibfnamefont{J.~M.} \bibnamefont{Campbell}},
  \bibinfo{author}{\bibfnamefont{W.~T.} \bibnamefont{Giele}}, \bibnamefont{and}
  \bibinfo{author}{\bibfnamefont{C.}~\bibnamefont{Williams}}
  (\bibinfo{year}{2012}{\natexlab{a}}), \eprint{1205.3434}.

\bibitem[{\citenamefont{Campbell
  et~al.}(2012{\natexlab{b}})\citenamefont{Campbell, Giele, and
  Williams}}]{Campbell:2012cz}
\bibinfo{author}{\bibfnamefont{J.~M.} \bibnamefont{Campbell}},
  \bibinfo{author}{\bibfnamefont{W.~T.} \bibnamefont{Giele}}, \bibnamefont{and}
  \bibinfo{author}{\bibfnamefont{C.}~\bibnamefont{Williams}}
  (\bibinfo{year}{2012}{\natexlab{b}}), \eprint{1204.4424}.

\bibitem[{\citenamefont{Chatrchyan
  et~al.}(2012{\natexlab{b}})}]{Chatrchyan:2012sn}
\bibinfo{author}{\bibfnamefont{S.}~\bibnamefont{Chatrchyan}}
  \bibnamefont{et~al.} (\bibinfo{collaboration}{CMS Collaboration}),
  \bibinfo{journal}{JHEP} \textbf{\bibinfo{volume}{1204}}, \bibinfo{pages}{036}
  (\bibinfo{year}{2012}{\natexlab{b}}), \eprint{1202.1416}.

\bibitem[{\citenamefont{Chatrchyan
  et~al.}(2012{\natexlab{c}})}]{Chatrchyan:2012ufa}
\bibinfo{author}{\bibfnamefont{S.}~\bibnamefont{Chatrchyan}}
  \bibnamefont{et~al.} (\bibinfo{collaboration}{CMS Collaboration}),
  \bibinfo{journal}{Phys.Lett.} \textbf{\bibinfo{volume}{B716}},
  \bibinfo{pages}{30} (\bibinfo{year}{2012}{\natexlab{c}}), \eprint{1207.7235}.

\bibitem[{\citenamefont{Chatrchyan et~al.}(2013)}]{Chatrchyan:2012jja}
\bibinfo{author}{\bibfnamefont{S.}~\bibnamefont{Chatrchyan}}
  \bibnamefont{et~al.} (\bibinfo{collaboration}{CMS Collaboration}),
  \bibinfo{journal}{Phys.Rev.Lett.} \textbf{\bibinfo{volume}{110}},
  \bibinfo{pages}{081803} (\bibinfo{year}{2013}), \eprint{1212.6639}.

\bibitem[{\citenamefont{Modak et~al.}(2013)\citenamefont{Modak, Sahoo, Sinha,
  and Cheng}}]{Modak:2013sb}
\bibinfo{author}{\bibfnamefont{T.}~\bibnamefont{Modak}},
  \bibinfo{author}{\bibfnamefont{D.}~\bibnamefont{Sahoo}},
  \bibinfo{author}{\bibfnamefont{R.}~\bibnamefont{Sinha}}, \bibnamefont{and}
  \bibinfo{author}{\bibfnamefont{H.-Y.} \bibnamefont{Cheng}}
  (\bibinfo{year}{2013}), \eprint{1301.5404}.

\bibitem[{\citenamefont{Sun et~al.}(2013)\citenamefont{Sun, Wang, and
  Gao}}]{Sun:2013yra}
\bibinfo{author}{\bibfnamefont{Y.}~\bibnamefont{Sun}},
  \bibinfo{author}{\bibfnamefont{X.-F.} \bibnamefont{Wang}}, \bibnamefont{and}
  \bibinfo{author}{\bibfnamefont{D.-N.} \bibnamefont{Gao}}
  (\bibinfo{year}{2013}), \eprint{1309.4171}.

\bibitem[{\citenamefont{Gainer et~al.}(2013)\citenamefont{Gainer, Lykken,
  Matchev, Mrenna, and Park}}]{Gainer:2013rxa}
\bibinfo{author}{\bibfnamefont{J.~S.} \bibnamefont{Gainer}},
  \bibinfo{author}{\bibfnamefont{J.}~\bibnamefont{Lykken}},
  \bibinfo{author}{\bibfnamefont{K.~T.} \bibnamefont{Matchev}},
  \bibinfo{author}{\bibfnamefont{S.}~\bibnamefont{Mrenna}}, \bibnamefont{and}
  \bibinfo{author}{\bibfnamefont{M.}~\bibnamefont{Park}},
  \bibinfo{journal}{Phys.Rev.Lett.} \textbf{\bibinfo{volume}{111}},
  \bibinfo{pages}{041801} (\bibinfo{year}{2013}), \eprint{1304.4936}.

\bibitem[{\citenamefont{Artoisenet
  et~al.}(2013{\natexlab{a}})\citenamefont{Artoisenet, de~Aquino, Demartin,
  Frederix, Frixione et~al.}}]{Artoisenet:2013puc}
\bibinfo{author}{\bibfnamefont{P.}~\bibnamefont{Artoisenet}},
  \bibinfo{author}{\bibfnamefont{P.}~\bibnamefont{de~Aquino}},
  \bibinfo{author}{\bibfnamefont{F.}~\bibnamefont{Demartin}},
  \bibinfo{author}{\bibfnamefont{R.}~\bibnamefont{Frederix}},
  \bibinfo{author}{\bibfnamefont{S.}~\bibnamefont{Frixione}},
  \bibnamefont{et~al.} (\bibinfo{year}{2013}{\natexlab{a}}),
  \eprint{1306.6464}.

\bibitem[{\citenamefont{Anderson et~al.}(2013)\citenamefont{Anderson,
  Bolognesi, Caola, Gao, Gritsan et~al.}}]{Anderson:2013fba}
\bibinfo{author}{\bibfnamefont{I.}~\bibnamefont{Anderson}},
  \bibinfo{author}{\bibfnamefont{S.}~\bibnamefont{Bolognesi}},
  \bibinfo{author}{\bibfnamefont{F.}~\bibnamefont{Caola}},
  \bibinfo{author}{\bibfnamefont{Y.}~\bibnamefont{Gao}},
  \bibinfo{author}{\bibfnamefont{A.~V.} \bibnamefont{Gritsan}},
  \bibnamefont{et~al.} (\bibinfo{year}{2013}), \eprint{1309.4819}.

\bibitem[{\citenamefont{Chen et~al.}(2013{\natexlab{b}})\citenamefont{Chen,
  Cheng, Gainer, Korytov, Matchev et~al.}}]{Chen:2013waa}
\bibinfo{author}{\bibfnamefont{M.}~\bibnamefont{Chen}},
  \bibinfo{author}{\bibfnamefont{T.}~\bibnamefont{Cheng}},
  \bibinfo{author}{\bibfnamefont{J.~S.} \bibnamefont{Gainer}},
  \bibinfo{author}{\bibfnamefont{A.}~\bibnamefont{Korytov}},
  \bibinfo{author}{\bibfnamefont{K.~T.} \bibnamefont{Matchev}},
  \bibnamefont{et~al.} (\bibinfo{year}{2013}{\natexlab{b}}),
  \eprint{1310.1397}.

\bibitem[{\citenamefont{Buchalla et~al.}(2013)\citenamefont{Buchalla, Cata, and
  D'Ambrosio}}]{Buchalla:2013mpa}
\bibinfo{author}{\bibfnamefont{G.}~\bibnamefont{Buchalla}},
  \bibinfo{author}{\bibfnamefont{O.}~\bibnamefont{Cata}}, \bibnamefont{and}
  \bibinfo{author}{\bibfnamefont{G.}~\bibnamefont{D'Ambrosio}}
  (\bibinfo{year}{2013}), \eprint{1310.2574}.

\bibitem[{\citenamefont{Chen and Vega-Morales}(2014)}]{Chen:2013ejz}
\bibinfo{author}{\bibfnamefont{Y.}~\bibnamefont{Chen}} \bibnamefont{and}
  \bibinfo{author}{\bibfnamefont{R.}~\bibnamefont{Vega-Morales}},
  \bibinfo{journal}{JHEP} \textbf{\bibinfo{volume}{1404}}, \bibinfo{pages}{057}
  (\bibinfo{year}{2014}), \eprint{1310.2893}.

\bibitem[{\citenamefont{Gainer et~al.}(2014)\citenamefont{Gainer, Lykken,
  Matchev, Mrenna, and Park}}]{Gainer:2014hha}
\bibinfo{author}{\bibfnamefont{J.~S.} \bibnamefont{Gainer}},
  \bibinfo{author}{\bibfnamefont{J.}~\bibnamefont{Lykken}},
  \bibinfo{author}{\bibfnamefont{K.~T.} \bibnamefont{Matchev}},
  \bibinfo{author}{\bibfnamefont{S.}~\bibnamefont{Mrenna}}, \bibnamefont{and}
  \bibinfo{author}{\bibfnamefont{M.}~\bibnamefont{Park}}
  (\bibinfo{year}{2014}), \eprint{1403.4951}.

\bibitem[{\citenamefont{Chen et~al.}(2014{\natexlab{a}})\citenamefont{Chen,
  Harnik, and Vega-Morales}}]{Chen:2014gka}
\bibinfo{author}{\bibfnamefont{Y.}~\bibnamefont{Chen}},
  \bibinfo{author}{\bibfnamefont{R.}~\bibnamefont{Harnik}}, \bibnamefont{and}
  \bibinfo{author}{\bibfnamefont{R.}~\bibnamefont{Vega-Morales}},
  \bibinfo{journal}{Phys.Rev.Lett.} \textbf{\bibinfo{volume}{113}},
  \bibinfo{pages}{191801} (\bibinfo{year}{2014}{\natexlab{a}}),
  \eprint{1404.1336}.

\bibitem[{CMS(2014)}]{CMS-PAS-HIG-14-014}
\bibinfo{type}{Tech. Rep.} \bibinfo{number}{CMS-PAS-HIG-14-014},
  \bibinfo{institution}{CERN}, \bibinfo{address}{Geneva}
  (\bibinfo{year}{2014}), \eprint{CMS Collaboration}.

\bibitem[{\citenamefont{Khachatryan et~al.}(2014)}]{Khachatryan:2014kca}
\bibinfo{author}{\bibfnamefont{V.}~\bibnamefont{Khachatryan}}
  \bibnamefont{et~al.} (\bibinfo{collaboration}{CMS Collaboration})
  (\bibinfo{year}{2014}), \eprint{1411.3441}.

\bibitem[{\citenamefont{Chen et~al.}(2014{\natexlab{b}})\citenamefont{Chen,
  Di~Marco, Lykken, Spiropulu, Vega-Morales et~al.}}]{Chen:2014hqs}
\bibinfo{author}{\bibfnamefont{Y.}~\bibnamefont{Chen}},
  \bibinfo{author}{\bibfnamefont{E.}~\bibnamefont{Di~Marco}},
  \bibinfo{author}{\bibfnamefont{J.}~\bibnamefont{Lykken}},
  \bibinfo{author}{\bibfnamefont{M.}~\bibnamefont{Spiropulu}},
  \bibinfo{author}{\bibfnamefont{R.}~\bibnamefont{Vega-Morales}},
  \bibnamefont{et~al.} (\bibinfo{year}{2014}{\natexlab{b}}),
  \eprint{1410.4817}.

\bibitem[{\citenamefont{Chen}(2014)}]{Thesis}
\bibinfo{author}{\bibfnamefont{Y.}~\bibnamefont{Chen}} (\bibinfo{year}{2014}),
  \eprint{Precision measurement of the $125$~GeV Higgs boson discovered in
  proton-proton collisions at $\sqrt{s} = 7,8$~TeV with the CMS detector at the
  LHC, Thesis in preparation for P.h.D. at California Institute of Technology}.

\bibitem[{\citenamefont{Y.Gao}(2013)}]{gaotalk}
\bibinfo{author}{\bibnamefont{Y.Gao}} (\bibinfo{year}{2013}), \eprint{Talk
  given at ATLAS Higgs WG Workshop,
  https://indico.cern.ch/event/280998/session\\
  /5/contribution/29/material/slides/0.pdf}.

\bibitem[{\citenamefont{Vega-Morales}(2013)}]{Vega-Morales:2013dda}
\bibinfo{author}{\bibfnamefont{R.}~\bibnamefont{Vega-Morales}}
  (\bibinfo{year}{2013}), \eprint{{The Higgs Boson as a Window to Beyond the
  Standard Model, Thesis in fulfillment of P.h.D requirements, Northwestern
  University.}}

\bibitem[{\citenamefont{Bradie}(2006)}]{Bradie:943117}
\bibinfo{author}{\bibfnamefont{B.}~\bibnamefont{Bradie}},
  \emph{\bibinfo{title}{{A friendly introduction to numerical analysis: with C
  and MATLAB materials on website; int. ed.}}}
  (\bibinfo{publisher}{Prentice-Hall}, \bibinfo{address}{Englewood Cliffs, NJ},
  \bibinfo{year}{2006}).

\bibitem[{\citenamefont{Artoisenet and Mattelaer}(2008)}]{Artoisenet:2008zz}
\bibinfo{author}{\bibfnamefont{P.}~\bibnamefont{Artoisenet}} \bibnamefont{and}
  \bibinfo{author}{\bibfnamefont{O.}~\bibnamefont{Mattelaer}},
  \bibinfo{journal}{PoS} \textbf{\bibinfo{volume}{CHARGED2008}},
  \bibinfo{pages}{025} (\bibinfo{year}{2008}).

\bibitem[{\citenamefont{Artoisenet et~al.}(2010)\citenamefont{Artoisenet,
  Lemaitre, Maltoni, and Mattelaer}}]{Artoisenet:2010cn}
\bibinfo{author}{\bibfnamefont{P.}~\bibnamefont{Artoisenet}},
  \bibinfo{author}{\bibfnamefont{V.}~\bibnamefont{Lemaitre}},
  \bibinfo{author}{\bibfnamefont{F.}~\bibnamefont{Maltoni}}, \bibnamefont{and}
  \bibinfo{author}{\bibfnamefont{O.}~\bibnamefont{Mattelaer}},
  \bibinfo{journal}{JHEP} \textbf{\bibinfo{volume}{1012}}, \bibinfo{pages}{068}
  (\bibinfo{year}{2010}), \eprint{1007.3300}.

\bibitem[{\citenamefont{Artoisenet
  et~al.}(2013{\natexlab{b}})\citenamefont{Artoisenet, de~Aquino, Maltoni, and
  Mattelaer}}]{Artoisenet:2013vfa}
\bibinfo{author}{\bibfnamefont{P.}~\bibnamefont{Artoisenet}},
  \bibinfo{author}{\bibfnamefont{P.}~\bibnamefont{de~Aquino}},
  \bibinfo{author}{\bibfnamefont{F.}~\bibnamefont{Maltoni}}, \bibnamefont{and}
  \bibinfo{author}{\bibfnamefont{O.}~\bibnamefont{Mattelaer}}
  (\bibinfo{year}{2013}{\natexlab{b}}), \eprint{1304.6414}.

\bibitem[{\citenamefont{Falkowski and Vega-Morales}(2014)}]{Falkowski:2014ffa}
\bibinfo{author}{\bibfnamefont{A.}~\bibnamefont{Falkowski}} \bibnamefont{and}
  \bibinfo{author}{\bibfnamefont{R.}~\bibnamefont{Vega-Morales}}
  (\bibinfo{year}{2014}), \eprint{1405.1095}.

\bibitem[{\citenamefont{Bredenstein
  et~al.}(2006{\natexlab{a}})\citenamefont{Bredenstein, Denner, Dittmaier, and
  Weber}}]{Bredenstein:2006rh}
\bibinfo{author}{\bibfnamefont{A.}~\bibnamefont{Bredenstein}},
  \bibinfo{author}{\bibfnamefont{A.}~\bibnamefont{Denner}},
  \bibinfo{author}{\bibfnamefont{S.}~\bibnamefont{Dittmaier}},
  \bibnamefont{and} \bibinfo{author}{\bibfnamefont{M.}~\bibnamefont{Weber}},
  \bibinfo{journal}{Phys.Rev.} \textbf{\bibinfo{volume}{D74}},
  \bibinfo{pages}{013004} (\bibinfo{year}{2006}{\natexlab{a}}),
  \eprint{hep-ph/0604011}.

\bibitem[{\citenamefont{Bredenstein
  et~al.}(2006{\natexlab{b}})\citenamefont{Bredenstein, Denner, Dittmaier, and
  Weber}}]{Bredenstein:2006nk}
\bibinfo{author}{\bibfnamefont{A.}~\bibnamefont{Bredenstein}},
  \bibinfo{author}{\bibfnamefont{A.}~\bibnamefont{Denner}},
  \bibinfo{author}{\bibfnamefont{S.}~\bibnamefont{Dittmaier}},
  \bibnamefont{and} \bibinfo{author}{\bibfnamefont{M.}~\bibnamefont{Weber}},
  \bibinfo{journal}{Nucl.Phys.Proc.Suppl.} \textbf{\bibinfo{volume}{160}},
  \bibinfo{pages}{131} (\bibinfo{year}{2006}{\natexlab{b}}),
  \eprint{hep-ph/0607060}.

\bibitem[{\citenamefont{Adam et~al.}(2008)\citenamefont{Adam, Aziz, Andersen,
  Belyaev, Binoth et~al.}}]{Adam:2008aa}
\bibinfo{author}{\bibfnamefont{N.}~\bibnamefont{Adam}},
  \bibinfo{author}{\bibfnamefont{T.}~\bibnamefont{Aziz}},
  \bibinfo{author}{\bibfnamefont{J.}~\bibnamefont{Andersen}},
  \bibinfo{author}{\bibfnamefont{A.}~\bibnamefont{Belyaev}},
  \bibinfo{author}{\bibfnamefont{T.}~\bibnamefont{Binoth}},
  \bibnamefont{et~al.}, pp. \bibinfo{pages}{215--289} (\bibinfo{year}{2008}),
  \eprint{0803.1154}.

\bibitem[{\citenamefont{Neyman and Pearson}(1933)}]{Neyman}
\bibinfo{author}{\bibfnamefont{J.}~\bibnamefont{Neyman}} \bibnamefont{and}
  \bibinfo{author}{\bibfnamefont{E.~S.} \bibnamefont{Pearson}}
  (\bibinfo{year}{1933}), \eprint{Philosophical Transactions of the Royal
  Society of London. Series A Vol. 231, pp. 289-337}.

\bibitem[{\citenamefont{CMS}(2006)}]{CMSperformance2}
\bibinfo{author}{\bibnamefont{CMS}} (\bibinfo{year}{2006}), \eprint{CMS
  Physics: Technical Design Report\\ Volume 1: Detector Performance and
  Software}.

\bibitem[{\citenamefont{CMS}(2013{\natexlab{a}})}]{CMSperformance}
\bibinfo{author}{\bibnamefont{CMS}} (\bibinfo{year}{2013}{\natexlab{a}}),
  \eprint{https://twiki.cern.ch/twiki/bin/view/CMS\\
  Public/EGMElectronsMoriond2013}.

\bibitem[{\citenamefont{CMS}(2013{\natexlab{b}})}]{CMS-DP-2013-003}
\bibinfo{author}{\bibnamefont{CMS}} (\bibinfo{year}{2013}{\natexlab{b}}),
  \eprint{Electron performance with 19.6 fb$^{-1}$ of data collected at
  $\sqrt{s} = 8$ TeV with the CMS detector}.

\bibitem[{\citenamefont{Chen and Vega-Morales}(2013)}]{WEBSITE}
\bibinfo{author}{\bibfnamefont{Y.}~\bibnamefont{Chen}} \bibnamefont{and}
  \bibinfo{author}{\bibfnamefont{R.}~\bibnamefont{Vega-Morales}}
  (\bibinfo{year}{2013}), \eprint{{Website under
  construction:http://yichen.me/project/GoldenChannel/}}.

\bibitem[{\citenamefont{Christensen and Duhr}(2009)}]{Christensen:2008py}
\bibinfo{author}{\bibfnamefont{N.~D.} \bibnamefont{Christensen}}
  \bibnamefont{and} \bibinfo{author}{\bibfnamefont{C.}~\bibnamefont{Duhr}},
  \bibinfo{journal}{Comput.Phys.Commun.} \textbf{\bibinfo{volume}{180}},
  \bibinfo{pages}{1614} (\bibinfo{year}{2009}), \eprint{0806.4194}.

\bibitem[{\citenamefont{Alwall et~al.}(2007)\citenamefont{Alwall, Demin,
  de~Visscher, Frederix, Herquet et~al.}}]{Alwall:2007st}
\bibinfo{author}{\bibfnamefont{J.}~\bibnamefont{Alwall}},
  \bibinfo{author}{\bibfnamefont{P.}~\bibnamefont{Demin}},
  \bibinfo{author}{\bibfnamefont{S.}~\bibnamefont{de~Visscher}},
  \bibinfo{author}{\bibfnamefont{R.}~\bibnamefont{Frederix}},
  \bibinfo{author}{\bibfnamefont{M.}~\bibnamefont{Herquet}},
  \bibnamefont{et~al.}, \bibinfo{journal}{JHEP}
  \textbf{\bibinfo{volume}{0709}}, \bibinfo{pages}{028} (\bibinfo{year}{2007}),
  \eprint{0706.2334}.

\bibitem[{\citenamefont{Chen et~al.}(2014{\natexlab{c}})\citenamefont{Chen,
  Stolarski, and Vega-Morales}}]{loopstudy}
\bibinfo{author}{\bibfnamefont{Y.}~\bibnamefont{Chen}},
  \bibinfo{author}{\bibfnamefont{D.}~\bibnamefont{Stolarski}},
  \bibnamefont{and}
  \bibinfo{author}{\bibfnamefont{R.}~\bibnamefont{Vega-Morales}}
  (\bibinfo{year}{2014}{\natexlab{c}}), \eprint{Work in preparation}.

\bibitem[{\citenamefont{Falkowski et~al.}(2013)\citenamefont{Falkowski, Riva,
  and Urbano}}]{Falkowski:2013dza}
\bibinfo{author}{\bibfnamefont{A.}~\bibnamefont{Falkowski}},
  \bibinfo{author}{\bibfnamefont{F.}~\bibnamefont{Riva}}, \bibnamefont{and}
  \bibinfo{author}{\bibfnamefont{A.}~\bibnamefont{Urbano}}
  (\bibinfo{year}{2013}), \eprint{1303.1812}.

\bibitem[{\citenamefont{Alwall et~al.}(2011)\citenamefont{Alwall, Herquet,
  Maltoni, Mattelaer, and Stelzer}}]{Alwall:2011uj}
\bibinfo{author}{\bibfnamefont{J.}~\bibnamefont{Alwall}},
  \bibinfo{author}{\bibfnamefont{M.}~\bibnamefont{Herquet}},
  \bibinfo{author}{\bibfnamefont{F.}~\bibnamefont{Maltoni}},
  \bibinfo{author}{\bibfnamefont{O.}~\bibnamefont{Mattelaer}},
  \bibnamefont{and} \bibinfo{author}{\bibfnamefont{T.}~\bibnamefont{Stelzer}},
  \bibinfo{journal}{JHEP} \textbf{\bibinfo{volume}{1106}}, \bibinfo{pages}{128}
  (\bibinfo{year}{2011}), \eprint{1106.0522}.

\bibitem[{\citenamefont{Melia et~al.}(2011)\citenamefont{Melia, Nason, Rontsch,
  and Zanderighi}}]{Melia:2011tj}
\bibinfo{author}{\bibfnamefont{T.}~\bibnamefont{Melia}},
  \bibinfo{author}{\bibfnamefont{P.}~\bibnamefont{Nason}},
  \bibinfo{author}{\bibfnamefont{R.}~\bibnamefont{Rontsch}}, \bibnamefont{and}
  \bibinfo{author}{\bibfnamefont{G.}~\bibnamefont{Zanderighi}},
  \bibinfo{journal}{JHEP} \textbf{\bibinfo{volume}{1111}}, \bibinfo{pages}{078}
  (\bibinfo{year}{2011}), \eprint{1107.5051}.

\bibitem[{\citenamefont{James}(1994)}]{James:1994vla}
\bibinfo{author}{\bibfnamefont{F.}~\bibnamefont{James}} (\bibinfo{year}{1994}),
  \eprint{{MINUIT Function Minimization and Error Analysis: Reference Manual
  Version 94.1}}.

\end{thebibliography}

\end{document}